\documentclass[prb,amsmath,amssymb,graphicx,reprint]{revtex4-1}
\usepackage{graphicx}
\usepackage{dcolumn}
\usepackage{bm}
\usepackage{booktabs}
\usepackage{tabularx}
\newcommand{\gt}{$\gamma^{\prime\prime\prime}$}

\newcommand{\gFeN}{$\gamma^{\prime}\mathrm{-Fe_{4}N}$}
\newcommand{\eFeyN}{$\varepsilon-\mathrm{Fe_{3-z}N}$}

\begin{document}


\title{The origin of anomalous diffusion in iron mononitride thin films}

\author{Akhil Tayal$^1$, Mukul Gupta$^1$}\email{mgupta@csr.res.in/dr.mukul.gupta@gmail.com}\author{Ajay Gupta$^2$, P. R. Rajput$^3$, J. Stahn$^4$}
\address{$^1$UGC-DAE Consortium for Scientific Research, University
Campus, Khandwa Road, Indore 452 001,India}
\address{$^2$Amity Center for Spintronic Materials, Amity University, Sector 125, Noida-201
303, India}
\address{$^3$Atomic \& Molecular Physics Division, Bhabha Atomic Research Centre, Mumbai 400085, India}
\address{$^4$Laboratory for Neutron Scattering and Imaging, Paul Scherrer Institute, CH-5232 Villigen, Switzerland}


\begin{abstract}

We have studied the origin of a counter intuitive diffusion
behavior of Fe and N atoms in a iron mononitride (FeN) thin film.
It was observed that in-spite of a larger atomic size, Fe tend to
diffuse more rapidly than smaller N atoms. This only happens in
the N-rich region of Fe-N phase diagram, in the N-poor regions, N
diffusion coefficient is orders of magnitude larger than Fe.
Detailed self-diffusion measurements performed in FeN thin films
reveal that the diffusion mechanism of Fe and N is different - Fe
atoms diffuse through a complex process, which in addition to a
volume diffusion, pre-dominantly controlled by a fast grain
boundary diffusion. On the other hand N atoms diffuse through a
classical volume-type diffusion process. Observed results have
been explained in terms of stronger Fe-N (than Fe-Fe) bonds
generally predicted theoretically for mononitride compositions of
transition metals.

\end{abstract}

\date{\today}
\maketitle




\section{Introduction}
\label{1}

Transition metal nitrides (TMN), specially 3$d$ TM mononitrides
(TMMN) are an important class of materials exhibiting several
interesting properties such as
superhardness~\cite{Veprek:Hard:99,PRL:Jhi:Vacancy:TMNs:01,Nature:Jhi:TMNs,Hao:PRL:superhard},
superconductivity~\cite{TMNs:superconductivity}, corrosion and
wear resistance~\cite{PRB:Steneteg:13:TMNs,Science:Sproul:hard},
etc. Along the 3$d$ series, there is a characteristic variation in
the heat of formation ($\Delta H_f^{\circ}$) for the 3$d$ TMMN -
an initial increases to a maximum is followed by a decrease and a
plateau in $\Delta H_f^{\circ}$.~\cite{PRB:1993:Haguland} This
inherently makes formation of early 3$d$ TMMN viz. ScN, TiN, VN,
CrN easier than that of late ones viz. MnN, FeN, CoN, NiN; e.g.
$\Delta H_f^{\circ}$=-\,338\,kJmol$^{-1}$ for TiN and
-\,47\,kJmol$^{-1}$ for FeN at 298\,K. As a result, mononitrides
of Ti,V,Cr can be easily prepared and possess excellent thermal
stability due to which they have been intensely
investigated.~\cite{PRB:Tsetseris:Ndefects:07,PRL:TMNs:07:Ndefects,Hultman:2000,Zhang:TMNs:2003:SCT,TiNZrNCrN:SCT:1998}
On the other hand, magnetic mononitrides (e.g. MnN,FeN, CoN)
started to gain attention rather
recently.~\cite{FeN:PRB:Houari,MG:JAC:2011,gupta:JAP2011,Jouanny2010TSF,Liu:CoN:14:JAC,Navio.PRB08,Wang:CoN:TSF:09,Bhattacharyya:FeN:Review}
Unlike early 3$d$ TMMN, FeN or CoN can only be formed in the form
of thin films using non-equilibrium processes such as reactive
sputtering~\cite{Schaaf.PMS.2002,gupta:JAP2011,MG:JAC:2011,JVSTA:Fang:CoN},
pulsed laser deposition~\cite{Gupta_JAC01}, and more recently by
molecular beam epitaxy assisted with a rf-discharge
nitrogen/ammonia source.~\cite{Naito:FeN:14,Navio.PRB08} Nickel
mononitrides are yet to be evidenced
experimentally.~\cite{Vempaire:Ni3N:JAP:09,Nishihara:Ni2N:2014} On
the basis of energetics of mononitrides, FeN and CoN are expected
to be metastable.

The metastable nature of FeN turned out to be a boon as FeN films
were exploited as a source of spin-injection to semiconductors or
diluted magnetic semiconductor in
spintronics.~\cite{Navio:APL:2009} FeN when heated above 650\,K
yields a thermally stable \gFeN~giving rise to an array of
lithographically defined spin-valves.~\cite{Navio:APL:2009} The
mechanism leading to such structural transformation was assumed to
be controlled by N-diffusion. XPS measurements for Fe 2$p$ and N
1$s$ peaks were used to measure N diffusion in
Fe.~\cite{Navio.NJP2010} However, the conclusion drawn from such
measurements that fast N diffusion leads to such transformations
can be misleading since interdiffusion of N in Fe, $not$
self-diffusion of N was measured.

Recent Fe and N self-diffusion measurements performed using
neutron reflectivity (NR) show that N self-diffusion is slower
than Fe.~\cite{gupta:JAP2011} This is a counter-intuitive result,
defying established diffusion models for binary metal-metalloid
systems where a smaller atom always diffuses faster than a larger
atom.~\cite{Faupel_RMP03} In absence of a suitable radioactive
tracer, N self-diffusion is rather difficult to
measure.~\cite{Hultman:2000,matzke1992diffusion} Nevertheless by
using $^{15}$N labelling, it can be obtained by doing
depth-profile measurements using secondary ion mass spectroscopy
(SIMS)~\cite{Harald.APL04,PRB:AT:2014}, nuclear reaction analysis
(NRA)~\cite{Hultman:2000} and NR. Among these, later provides an
unique opportunity to measure self-diffusion lengths down to
0.1\,nm and a possibility to measure N self-diffusion in the low
temperature regime (below
500\,K).~\cite{Harald.PRL06,gupta:JAP2011}

In the present work, we have carried out a study of Fe and N
self-diffusion process to understand the origin of anomalous
diffusion in non-magnetic FeN compound. Thin film samples were
prepared using nitrogen alone as a sputtering gas in a reactive
magnetron sputtering process. X-ray diffraction, absorption and
M\"{o}ssbauer spectroscopy measurements confirmed ZnS-type
structure of samples. NR measurements show that N diffusion is
slower than Fe, however detailed diffusion mechanism was obtained
from SIMS depth profiles using Le Claire's
analysis~\cite{LeClaire,Book:Kaur1995} for grain-boundary ($gb$)
diffusion. It was found that in the low temperatures regime (up to
550\,K) fast Fe diffusion takes place predominantly through the
$gb$ regions while N diffusion is a conventional volume type
diffusion. As the temperature is increased beyond it, the
difference between Fe and N diffusion decreases leading to
structural transformations - essentially triggered by Fe diffusion
and followed by N diffusion to the extent that N diffuses out of
the system. This is an important result defying a general
misconception the fast N diffusion leads to structural instability
and can be applied to understand the thermal stability of
transition metal mononitrides.

\section{Experimental}
\label{2}

In a direct current-magnetron sputtering (dc-MS) technique an iron
target (purity 99.95\%) was sputtered using nitrogen (purity
99.9995\%) alone as the sputtering medium at a constant power of
100\,W. N$_2$ gas was flown at a constant flow of 10\,sccm
yielding a pressure of about 0.4\,Pa, while the pressure before
gas flow was about 1$\times$10$^{-5}$\,Pa. Fe target was
pre-sputtered for 10\,minutes using Ar gas to remove surface
contaminations. Following samples were prepared at ambient
temperature on Si(100) and float glass substrates:\\
(N1):~[FeN(7.5\,nm)$\mid ^{57}$FeN(7.5\,nm)]$_{\times 10}$\\
(N2):~[FeN(7.5\,nm)$\mid$Fe$^{15}$N(7.5\,nm)]$_{\times 10}$\\
(S):~~~[FeN(100\,nm)$\mid
^{57}$Fe$^{15}$N(2\,nm)$\mid$FeN(100\,nm)]

Samples (N1) and (N2) were used to measure Fe and N self-diffusion
using NR, sample (S) was used to measure Fe and N self-diffusion
\textit{simultaneously} using SIMS. Natural Fe and $^{57}$Fe
($\sim$95\% enriched) targets were sputtered using natural
nitrogen and $^{15}$N ($\sim$98\% enriched) gases. To avoid any
mixing of nitrogen isotope gases, chamber was evacuated after
deposition of each layer and gas flows were monitored using a
residual gas analyzer (RGA).

The long-range structure transformation in the samples were
studied using x-ray diffraction (XRD) using a standard
diffractometer (Bruker D8Advance) equipped with Cu\,K-$\alpha$
x-ray source and a silicon stripe detector (Lynxeye). The local
structure was investigated using x-ray absorption spectroscopy
(XAS) and conversion electron M\"{o}ssbauer spectroscopy (CEMS).
The XAS measurements at Fe K-edge were performed at BL-9
beamline~\cite{BL09:Indus2} and those at N K-edge at BL-1
beamline~\cite{Phase:SXAS:BL01}, both at Indus-2 synchrotron
radiation source at Indore. While Fe K-edge measurements were
carried out in fluorescence mode, those at N K-edge were measured
in total electron yield mode. To study the phase transformation,
samples were annealed using a vacuum furnace for about two hours
at each temperatures. NR measurements were performed at AMOR
reflectometer at SINQ/PSI, Villigen, Switzerland. SIMS
measurements were carried out on a Hiden Analytical SIMS
Workstation using O$^{+}_{2}$ as primary with 5\,keV energy and
400\,nA beam current. The base pressure in SIMS measurement
chamber was 1$\times 10^{-7}$\,Pa and during measurement pressure
was of the order of 1$\times 10^{-6}$\,Pa due to differentially
pumping of oxygen gas source.

\section{Results and discussion}
\label{RSandDis}

\subsection{Structural measurements}
\label{XRD}

\begin{figure} \center
\includegraphics [width=85mm,height=65mm] {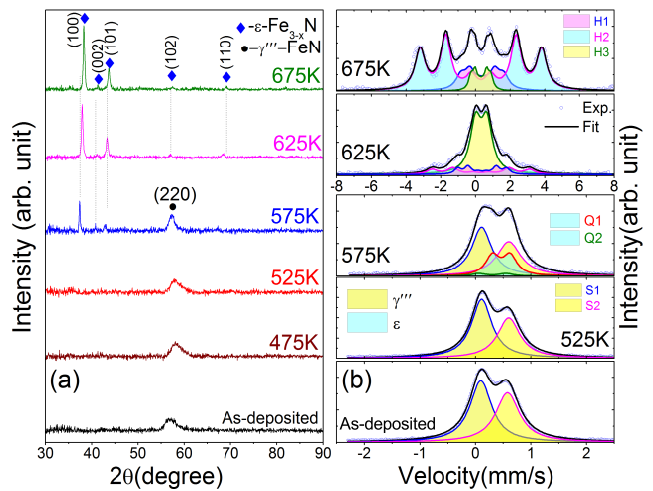}
\caption{\label{fig:XRDcems} XRD(a) and CEMS(b) patterns of sample
(N1) in the as-deposited state and after annealing at different
temperatures.}
\end{figure}

Figure~\ref{fig:XRDcems}(a) shows XRD patterns of sample (N1) in
the as-deposited state and after annealing at various
temperatures. In the as-deposited state, a single broad peak
appears around 2$\theta$=56 degree. From CEMS and XAS measurements
(shown later), it was confirmed that the formed phase is \gt-FeN
having ZnS-type structure oriented along (220)
plane.~\cite{Jouanny2010TSF,MG:JAC:2011} After annealing at
475\,K, a shift in the peak position towards higher angle was
observed. However, the XRD pattern of sample annealed at 525\,K is
almost identical to that of previously annealed sample. Such shift
in the peak position is an indication of annihilation of free
volume leading to densification of the
structure.~\cite{Guinier_XRD,warrenXRD} It is interesting to note
that the average crystallite size ($g_s$) (calculated using
Scherrer formula~\cite{Cullity_XRD}) remains at a value of about
$\sim5$\,nm upto an annealing temperature of 525\,K. At a higher
temperature of 575\,K, additional peaks corresponding to \eFeyN
(0$\leq$z$\leq$1) start to appear. With further increasing
annealing temperatures, growth of \eFeyN~phase can be seen. Along
with it a continuous shift of peaks position towards higher
2$\theta$ values were observed indicating `$z\rightarrow0$'.

Figure~\ref{fig:XRDcems}(b) shows CEMS spectra for sample (N1) in
the as-deposited state and after annealing. Obtained CEMS spectra
were fitted using a computer program
NORMOS/SITE~\cite{Normos:brand:95} and fitted parameters are given
in table~\ref{tab:MOSS}. As-deposited and the sample annealed at
525\,K shows an asymmetric doublet which is typically observed for
\gt-FeN in ZnS-type structure. Such spectra can be de-convoluted
into two singlets: one with a smaller and other with a larger
value of isomer shift ($\delta$). The singlet with
$\delta$=0.01\,mm/s corresponds to Fe surrounded tetrahedrally to
N, while other singlet originates due to defects and
vacancies.~\cite{Borsa.HI.2003,Jouanny2010TSF,MG:JAC:2011}
Annealing above 575\,K results in appreciable changes in CEMS
spectrum, at this temperature it can be best fitted using mixture
of two singlets and two quadrupole split doublets corresponding to
\gt~and \eFeyN~phase, respectively. Obtained fitting parameters
match-well with the reported
values.~\cite{Borsa.HI.2003,Schaaf.PMS.2002} It is known that
\eFeyN~phase exist in a wide composition range in which its
magnetic properties also get tuned with nitrogen
concentration.~\cite{JPC:chen:FexN:1983} It was observed that, at
room temperature, as `$z\rightarrow$1' \eFeyN~phase becomes
non-magnetic, whereas as `$z\rightarrow$0' it becomes
ferromagnetic. Appearance of sextet confirms ferromagnetic
ordering which is also supported by XRD results indicating
`$z\rightarrow$0' in \eFeyN. Obtained fitting parameters
(table~\ref{tab:MOSS}) for CEMS spectra measured above 575\,K
match-well with the reported values.~\cite{JPC:chen:FexN:1983}

\begin{table} \center
\caption{\label{tab:MOSS} Conversion electron M\"{o}ssbauer
spectroscopy (CEMS) parameters for iron mononitride thin films in
the as-deposited state (300\,K) and after annealing at various
temperatures. Here S stands for a singlet, Q for a doublet and H
for a sextet, $\delta$ for isomer shift, $\Gamma$ for quadrupole
splitting, $\mathbf{H}$ for hyperfine field and RA for relative
area.}\vspace{2mm}
\begin{tabularx}{\linewidth}{>{\centering\arraybackslash}X>{\centering\arraybackslash}X>{\centering\arraybackslash}X>{\centering\arraybackslash}X>{\centering\arraybackslash}X>{\centering\arraybackslash}X}
\hline \hline T&Component&$\delta$& $\Gamma$& $\mathbf{H}$ & RA \\
(K)&&(mm/s)&(mm/s)&(Tesla)&(\%) \\
&&$\pm0.03$&$\pm0.03$&$\pm0.2$&$\pm2$ \\
\hline \hline
300&S1&0.10&--&--&54\\
 &S2&0.57&--&--&46\\ \hline
525&S1&0.11&--&--&57\\
 &S2&0.60&--&--&43\\ \hline
575&S1&0.11&--&--&40\\
 &S2&0.60&--&--&34\\
&Q1&0.46&0.31&--&23\\
&Q2&0.30&0.50&--&3\\ \hline
625&H1&0.38&0.0&8.7&13\\
 &H2&0.31&0.0&17.4&17\\
 &H3&0.33&0.65&--&70\\ \hline
675&H1&0.40&0.0&7.6&22\\
 &H2&0.31&0.0&21.7&68\\
 &H3&0.30&0.0&3.6&10\\ \hline \hline
\end{tabularx}
\end{table}

\begin{figure} \center
\includegraphics [width=50mm,height=40mm] {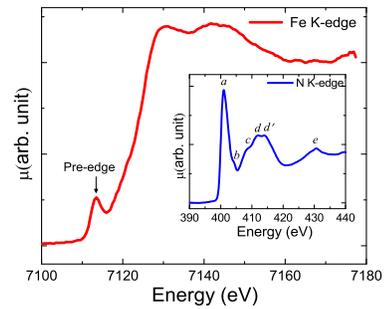}
\vspace {-1mm} \caption{\label{fig:XAS} Fe K-edge XAS spectra of
iron mononitride thin films. Inset of the figure shows spectra
taken at N K-edge.}
\end{figure}

To get precise information about the local structure of Fe and N,
XAS measurements at Fe and N K-edge were performed on sample (N1)
and are shown in figure~\ref{fig:XAS}. A strong pre-edge peak
around 7113\,eV can be seen before Fe K-edge. This is a signature
of quadrupole transition and its probability strongly depends on
the direction of the local electric field, which in turns gets
influenced by local site symmetry.~\cite{Bunker:XAFS} For such a
pre-edge peak to appear, inversion symmetry must be broken
(transition from a bound core level to a higher level empty
state). This is only possible when Fe is surrounded tetrahedrally
to N atoms because this arrangement is asymmetric under
inversion.~\cite{Bunker:XAFS} This clearly indicates ZnS-type
structure (tetrahedral coordination of Fe atoms) in our samples.
Such pre-edge feature has been frequently used to assign
tetrahedral or octahedral coordination in transition metal
complexes such as Mn~\cite{Pre-edge:XAS:Yamamoto},
Cr~\cite{Pre-edge:XAS:Cr}, etc. To further confirm this result,
XAS spectra was also taken at N K-edge (shown in the inset of
figure~\ref{fig:XAS}). Apart from edge feature `$a$' at 401\,eV,
three features that are assigned as ($c$, $d$ and $e$), are
observed due to the electronic transition to empty N-2p state
hybridized to Fe-3d state.~\cite{Jouanny2010TSF} Various TMMN
(viz. Ti, Cr, V) having NaCl type structure, the feature `$a$'
splits into two components due to crystal field splitting caused
by the octahedral coordination of N atoms surrounding the metal
ions.~\cite{SSC:Mitterbauer:EELS:04,Pfluger:EELS:89} Absence of
any splitting around the feature `$a$' is a clear indication of
tetrahedral coordination of Fe surrounding N atoms. Less intense
features ($b$) and ($d'$) observed in the spectra are due to
surface oxidation state.~\cite{Jouanny2010TSF} Combining the
information obtained from XAS measurements, it can be confirmed
that FeN phase has a ZnS-type structure. This is also in agreement
with low temperature high field M\"{o}ssbauer spectroscopy
measurements in this compound.~\cite{MG:JAC:2011}

\subsection{Self-diffusion measurements}
\begin{figure} \center
\includegraphics [width=80mm,height=45mm] {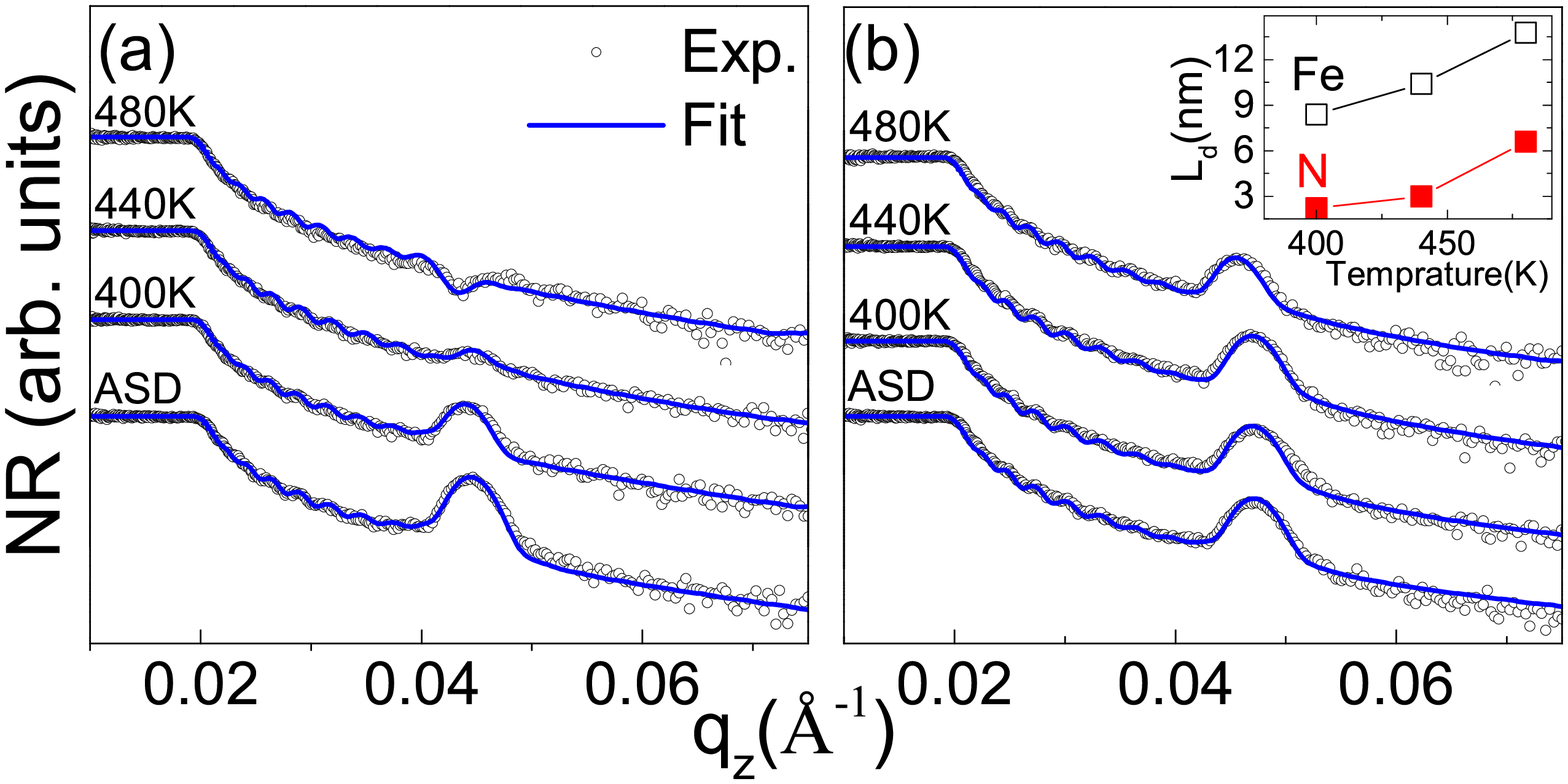}
\vspace {-5mm} \caption{\label{fig:NR} NR patterns of sample (N1)
(a) and (N2)(b) in the as-deposited state (ASD) and after
annealing at different temperatures for 2\,hours. Here scatter
points are experimental data and solid line is fit to them. Inset
of figure(b) shows variation of diffusion length
(L$_{\mathrm{d}}$) obtained by fitting. Here solid and open
symbols represent for Fe and N diffusion, respectively. Typical
error bars in calculating L$_{\mathrm{d}}$ are less than size of
symbols.}
\end{figure}

NR is a precise tool to measure atomic self-diffusion and
diffusivity as low as
1$\times$10$^{-25}\,\mathrm{m}^{2}\mathrm{s}^{-1}$ have been
measured using this
technique.~\cite{Spapen:APL80,Greer-JMMM96,gupta:PRB04,Harald.PRL06,PRB:AT:2014}
It is known that neutron scattering length ($b_n$) varies for
isotopes for $^{\mathrm{natural}}$Fe, $^{57}$Fe,
$^{\mathrm{natural}}$N, and $^{15}$N $b_n$=9.45\,fm, 2.3\,fm,
9.36\,fm and 6.6\,fm, respectively. Therefore, periodic isotope
multilayer are widely used to study atomic self-diffusion using
NR.~\cite{Spapen:APL80,Greer-JMMM96,gupta:PRB04,Harald.PRL06,PRB:AT:2014,gupta:JAP2011}
Figure~\ref{fig:NR} (a) and (b) shows NR patterns for samples (N1)
and (N2), respectively. Bragg peaks appearing due to isotopic
contrast of $^{\mathrm{nat}}$Fe/$^{57}$Fe and
$^{\mathrm{nat}}$N/$^{15}$N can be seen clearly. Patterns were
fitted using a computer program based on Parratt
formulism~\cite{Parratt.PR54,Parratt32} and obtained layer
thickness for sample (N1) is 7.9\,nm, (N2) is 7.3\,nm, close to
nominal values.

After annealing the intensity of Bragg peak start to decay and
such decay is more rapid for sample (N1) than for (N2). A decay of
Bragg peak intensity is a measure of atomic diffusion across
interfaces. Obtained results clearly indicate that Fe
self-diffusion is faster than N. Detailed fitting of NR data
yields diffusion length(L$_{\mathrm{d}}$), which are plotted in
the inset of figure~\ref{fig:NR}(b). Clearly, L$_{\mathrm{d}}$ is
significantly larger for Fe than for N. This result, although
counter intuitive, agrees well with previous studies on \gt-FeN
compound.~\cite{gupta:JAP2011} To get further insight leading to
such anomalous behavior, we did detailed SIMS measurements on
sample (S). With SIMS, unlike NR, both Fe and N self-diffusion can
be measured simultaneously.

\begin{figure} \center
\includegraphics [width=85mm,height=60mm] {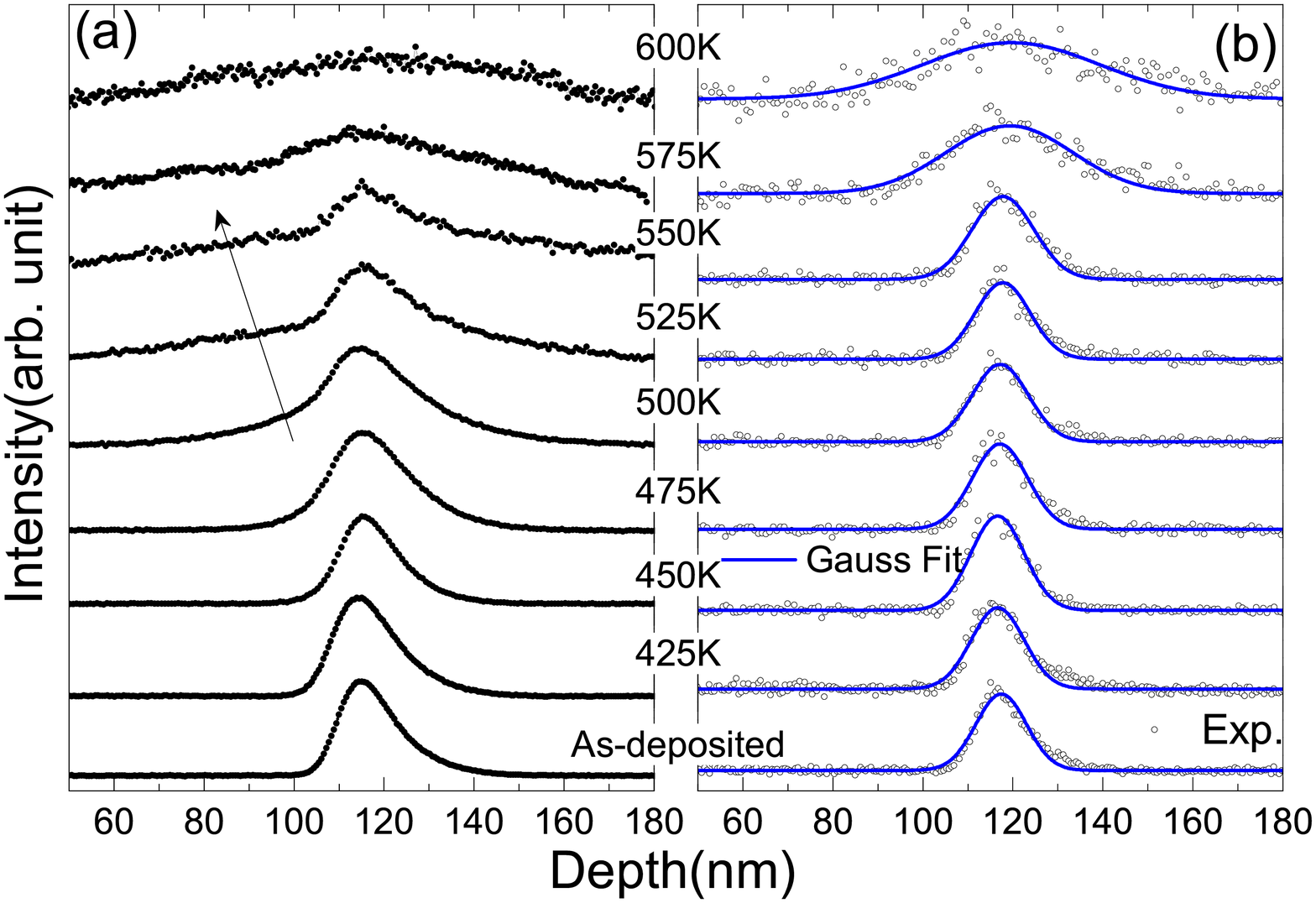}
\caption{\label{fig:isoSIMS} SIMS depth profile of $^{57}$Fe(a)
and $^{15}$N(b) for sample (S) after annealing at different
temperatures.}
\end{figure}

\begin{figure} \center
\includegraphics [width=85mm,height=40mm] {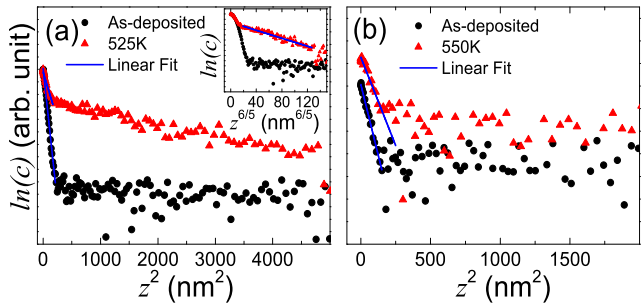}
\caption{\label{fig:FitSIMS} SIMS depth-profile for sample (S)
showing $^{57}$Fe(a) and $^{15}$N(b) distribution plotted versus
$z^2$ to calculate volume diffusion, in the as-deposited and
annealed state. Inset of (a) shows depth-profile of $^{57}$Fe
plotted versus $z^{6/5}$ to calculate grain-boundary diffusion.}
\end{figure}

Figure~\ref{fig:isoSIMS} shows SIMS depth-profile in sample (S)
for $^{57}$Fe(a) and $^{15}$N(b), isochronally annealed for 1\,hr
at each temperature. As annealing temperature is increased, the
width of $^{57}$Fe profiles become broader than those of $^{15}$N,
indicating faster Fe self-diffusion than N, in agreement with NR
results. For such SIMS depth-profiles, concentration profile of an
element at a depth $z$ can be fitted using a Gaussian distribution
function to obtain volume diffusivity($D_{V}$)
using~\cite{gupta:PRB02}:

\begin{equation}
\label{equ:SIMS} c(z,t)= \frac{\mathrm{const.}}{\sqrt{\pi
D_{V}t}}\mathrm{exp}\left(\frac{-z^{2}}{4D_{V}t}\right)
\end{equation}

\begin{figure} \center
\includegraphics [width=65mm,height=85mm] {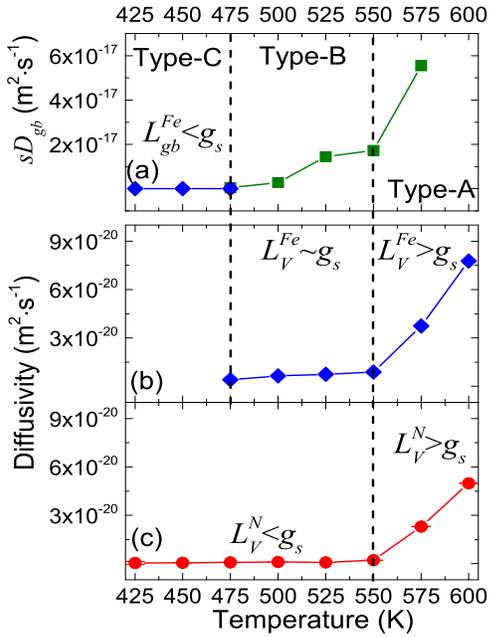}
\caption{\label{fig:Diffusivity} Variation of $D_{gb}$ of Fe(a),
$D_{V}$ of Fe(b) and $D_{V}$ of N(c) at different annealing
temperatures. Typical error bars in estimation of diffusivity are
of the order of size of symbols.}
\end{figure}

Here $t$ is annealing time at a temperature $T$. Alternatively, a
linear relation between $ln(c)$ and $z^2$, yields slope of the
fitted line, given by: slope=$-1/4D_{V}t$. Obtained N
diffusivities are shown in ~\ref{fig:Diffusivity}(c). While
$^{15}$N profiles fits well to this equation, $^{57}$Fe profiles
cannot be fitted using equation~\ref{equ:SIMS} alone, specially
above 475\,K. As shown more clearly in figure~\ref{fig:FitSIMS},
$^{15}$N profiles have single slope, $^{57}$Fe profiles have two
distinct slopes. Fe diffusivity upto 475\,K can be obtained using
equation~\ref{equ:SIMS}, above it, the additional slope can be
fitted using Le Claire's analysis~\cite{LeClaire} for $gb$
diffusion. An skewness in $^{57}$Fe depth-profiles above 475\,K
(marked by an arrow in figure~\ref{fig:isoSIMS}(a)) is an
indication of competing diffusion processes taking place through
grain-boundaries in addition to a volume type
diffusion.~\cite{Book:Kaur1995,GBD:Mishin:1999} Since
$D_{gb}$$>>$$D_{V}$, it can be assumed that fast Fe diffusion
observed in our samples is primarily $gb$ diffusion. In this case,
$D_{gb}$ can be obtained from the slope of `ln$c$' versus
`$z^{6/5}$' curve, using Suzuoka's instantaneous-source solution
equation~\cite{Suzuoka:GBD,Book:Kaur1995}:
\begin{equation}
\label{equ:GBD} s\delta D_{gb}= 1.084
\left(\frac{D^{0.91}_{V}}{t^{1.03}}\right)^{1/1.94}
\left(-\frac{\partial \mathrm{ln}c}{\partial
z^{6/5}}\right)^{-5/2.91}
\end{equation}

Here, $s$ is segregation factor, $\delta$ is width of $gb$. As
such experimental estimation of $\delta$ is difficult, a good
approximation is $\delta\sim$0.5\,nm, as often found in
literature.~\cite{GB:Diffusion:Fisher} Eq.~\ref{equ:GBD} is
applicable for the condition in which a dimensionless quantity
$\beta = \frac{s \delta D_{gb}}{2D^{3/2}_V t^{1/2}}$, has values
between 10 and 100. Since in our case we find $\beta\sim 23$ at
500\,K, equation~\ref{equ:GBD} can be applied to calculate
$sD_{gb}$. Inset of the figure~\ref{fig:FitSIMS}(a) shows a linear
fit to a curve plotted between `ln$c$' and $z^{6/5}$' after
annealing at 525\,K. Similar analysis was carried out at other
annealing temperatures and obtained values of $sD_{gb}$ and
$D_{V}$ are shown figure~\ref{fig:Diffusivity} (a) and (b),
respectively.

Grain-boundary diffusion is a complex process in which several
competing processes may take place
simultaneously.~\cite{GBD:Mishin:1999} However for a given
temperature range and duration, only few of them may get
activated. Thus different kinetic regimes of $D_{gb}$ have been
observed, ascribed by Harrison.~\cite{Harrison,Book:Kaur1995}
According to this classification, $D_{gb}$ has three regimes known
as type-A, B, and C. These regimes can be differentiated according
to a relation between $gb$ penetration depth
($L_{gb}=\sqrt{(D_{gb}t)}$), volume penetration depth
($L_{V}=\sqrt{(D_{V}t)}$) and grain size
($g_s$).~\cite{Mishin:95:TypeABC} For type-C regime $L_{gb}<g_s$,
for type-B regime $L_{V}<g_s$, and for type-A regime
$L_{V}>g_s$~\cite{Mishin:95:TypeABC}. Using these inequalities,
obtained values $D_{V}$ and $D_{gb}$ can be divided into distinct
kinetic regimes as shown in figure~\ref{fig:Diffusivity}. We find
that for Fe diffusion, $L_{gb}<g_s$ ($L_{gb}\sim$2\,nm,
$g_s\sim$5\,nm) for T$<$475\,K; for 475\,K$<$T$<$550\,K,
$L_{V}\sim g_s$ and $L_{V}> g_s$ for T$>$550\,K corresponding to
type-C, B and A kinetics, respectively.

On the other hand for N diffusion, we find $L_{V}<g_s$ below
550\,K and above it $L_{V}>g_s$. This indicates N diffusion is
taking place within a grain below 550\,K and as temperature is
raised beyond it, multi-grain N diffusion takes over. It appears
that Fe and N diffusion mechanism is significantly different.
While only $D_{V}$ takes place for N, Fe diffusion process seems
to be more complex having distinct kinetic regimes C, B and A as
annealing temperature increases. It is known that in type-C
regime, diffusion takes place pre-dominantly through
grain-boundaries leading to segregating of Fe atoms in the $gb$
region. In type-B regime, in addition to $gb$ diffusion, $D_{V}$
starts. Though it is not too effective to cause any structural or
magnetic transformation in this case. Finally above 550\,K,
$D_{V}$ of both Fe and N becomes comparable. Observed diffusion
behavior gives a direct correlation between Fe and N
self-diffusion with phase transformation. This is contrary to
general preassumption that only N diffusion leads to phase
transformation.

\begin{figure} \center
\includegraphics [width=85mm,height=70mm] {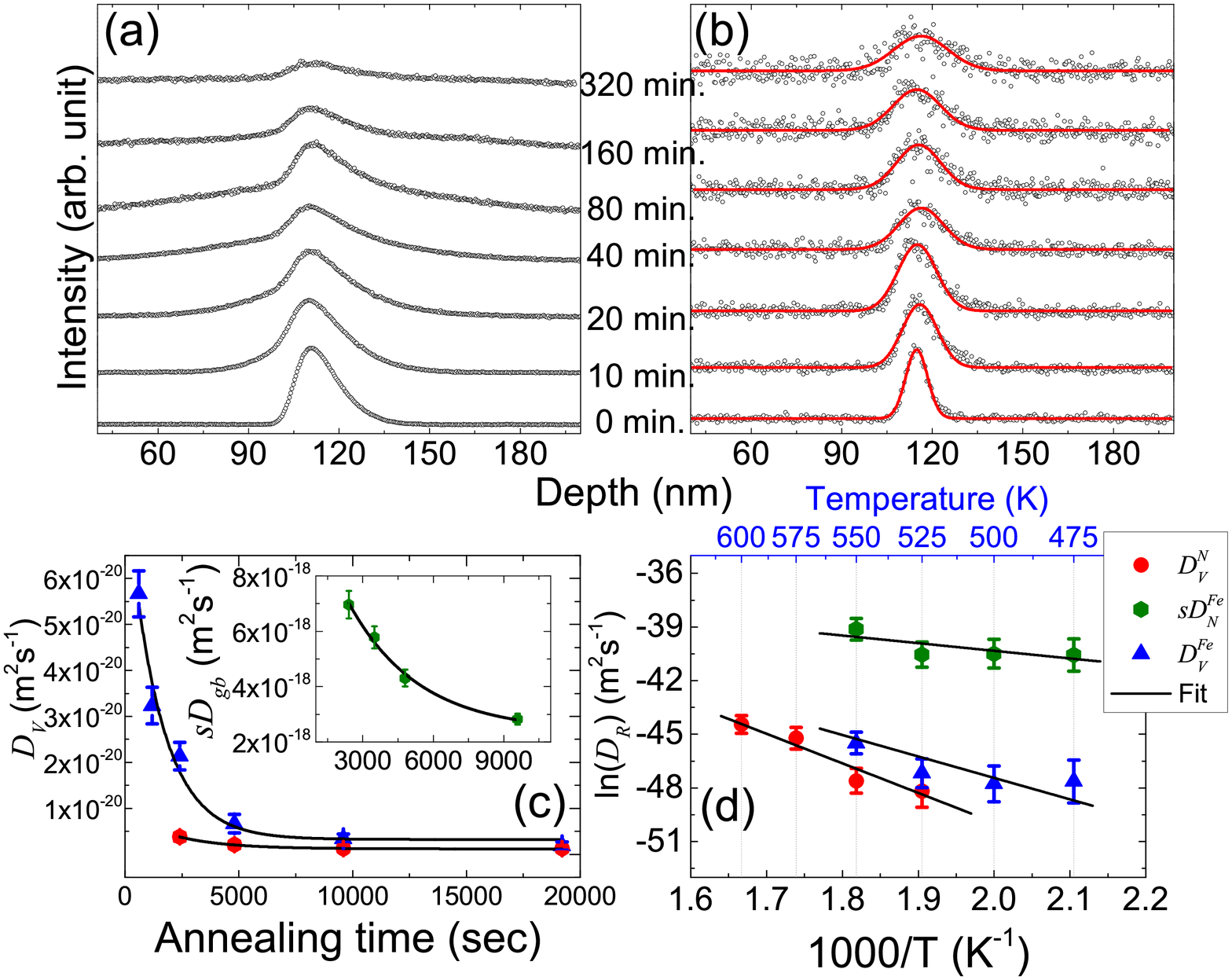}
\caption{\label{fig:aeplot} SIMS depth-profile of $^{57}$Fe(a) and
$^{15}$N(b) for sample (S) annealed at 525\,K for different
annealing times. Obtained values of $D_{V}$ for Fe and N annealed
for different times at 525\,K (c). Inset of figure (c) shows
variation in $D_{gb}$ of Fe. Arrhenius behavior of volume and
grain-boundary diffusion of Fe and N (d).}
\end{figure}

Since isochronal diffusion measurements only give a snap shot of
diffusion process, more insight about involved diffusion mechanism
was obtained by doing detailed isothermal diffusion measurements
between 475-550\,K (for Fe) and 525-600\,K (for N) in the steps of
25\,K for different annealing times. Representative SIMS
depth-profiles taken at 525\,K after various annealing times are
shown in figure~\ref{fig:aeplot} for Fe (a) and N (b). Following a
similar process D$_V$ and D$_{gb}$ was calculated and shown in
figure~\ref{fig:aeplot}(c). Time-dependent diffusivity data can be
fitted using:

\begin{equation}
\label{equ:relaxation} D = D_{R}+ A\cdot exp(-t/\tau)
\end{equation}

Where, $D_{R}$ is diffusivity in relaxed state, $A$ a constant,
$t$ annealing time, and $\tau$ is relaxation time. Using
~\ref{equ:relaxation} we get $\tau$ =1445($\pm$300)s, and
2192($\pm$300)s, respectively for Fe and N $D_{V}$; while for Fe
D$_{gb}$, $\tau$=2986($\pm$300)s. Much longer values of $\tau$ for
$gb$ diffusion indicate that it remain active for a longer time
due to availability of large volume of $gb$. Obtained values of
$D_{R}$ follows Arrhenius behavior given by:

\begin{equation}
\label{equ:Arrhenius} D_R = D_{0}exp(-E/k\mathrm{_B}T)
\end{equation}

Here, $D_0$ denotes pre-exponential factor, $E$ activation energy,
$T$ temperature and $k\mathrm{_B}$ Boltzmann's constant. The
obtained values of $E$ for $D_{V}$ of Fe and N are 1.0$\pm$0.2\,eV
and 1.4$\pm$0.2\,eV, respectively while for Fe $D_{gb}$ it is
0.6$\pm$0.2\,eV. As expected, smaller value of $E$ for the $gb$
diffusion signifies that it starts at a relatively lower
temperature as compared to $D_{V}$. Interestingly, we observe a
slightly higher value of the activation energy for $D_{V}$ of N as
compared to Fe (also $\tau$ for N $D_{V}$ was larger than Fe).
Observed discrepancy within $D_{V}$ of Fe and N can be understood
due to a stronger metal-nitrogen (than metal-metal) interaction
predicted theoretically for
TMMN~\cite{PRB:1993:Haguland,Paduani:CuNCoNNiN:2008}. Unlike
oxides, TMN are stabilized by a dominant metal-metal interaction
and therefore metal-metal bonds are stronger. However mononitride
having ZnS-type or NaCl-type structure, a volume expansion of
$fcc$ lattice takes place to accommodate N atoms in the
interstitial positions. Early theoretical calculations predicted
that the interaction distances are larger and bond energies are
significantly smaller for metal-metal bonds as compared to
metal-nitrogen bonds for 3$d$ TMMN.~\cite{Eck:JMC:99:TMNs} More
recent band structure calculations based of density-functional
theory, observed localization of metal valance bonds leading to
weakened metal-metal bonding in
mononitrides.~\cite{Paduani:CuNCoNNiN:2008} In additions a
suppression in N migration vacancy was predicted for slightly
off-stoichiometry mononitrides of various
TMN.~\cite{PRB:Tsetseris:Ndefects:07,PRL:TMNs:07:Ndefects} From
these theoretical studies it can be inferred that metal-metal
bonds are readily broken. This situation may lead to segregation
of Fe atoms in the $gb$ region, leading to faster Fe diffusion as
compared to N diffusion, under identical thermodynamic conditions.

\section{Conclusion}
\label{4} Iron mononitride thin films studied in this work, show
an anomalous self-diffusion behavior in which N atoms tend to
diffuse slower than Fe atoms. SIMS depth-profile measurements
reveal that the diffusion mechanism for Fe and N is different.
While N diffuses only via a volume-type diffusion process, Fe in
addition to volume, pre-dominantly diffuses through
grain-boundaries. Fe grain-boundary diffusion was about two orders
of magnitude more than its volume diffusion. Even for volume
diffusion, N diffusion was found to be less than Fe. This can be
understood in terms of stronger metal-nitrogen bonds (than
metal-metal) predicted theoretically for transition metal
mononitrides, evidenced experimentally in this work.

\section*{Acknowledgments}

A part of this work was performed at AMOR, Swiss Spallation
Neutron Source, Paul Scherrer Institute, Villigen, Switzerland. We
acknowledge D. M. Phase, D. K. Shukla, R. Sah and S. Karwal for
utilization of BL1 beamline and S. N. Jha, D. Bhattachrya for BL9
beamline. We are thankful to V.\,R.\,Reddy and A. Gome for CEMS
measurements; M. Horisberger for sample preparation; L. Behra for
XRD and SIMS measurements. We are thankful to A.\,K.\,Sinha and
V.\,Ganesan for support and encouragement. A.T. is thankful to
CSIR India for a research fellowship.


\begin{thebibliography}{10}%
\makeatletter
\providecommand \@ifxundefined [1]{%
 \ifx #1\undefined \expandafter \@firstoftwo
 \else \expandafter \@secondoftwo
\fi
}%
\providecommand \@ifnum [1]{%
 \ifnum #1\expandafter \@firstoftwo
 \else \expandafter \@secondoftwo
\fi
}%
\providecommand \enquote [1]{``#1''}%
\providecommand \bibnamefont  [1]{#1}%
\providecommand \bibfnamefont [1]{#1}%
\providecommand \citenamefont [1]{#1}%
\providecommand\href[0]{\@sanitize\@href}%
\providecommand\@href[1]{\endgroup\@@startlink{#1}\endgroup\@@href}%
\providecommand\@@href[1]{#1\@@endlink}%
\providecommand \@sanitize [0]{\begingroup\catcode`\&12\catcode`\#12\relax}%
\@ifxundefined \pdfoutput {\@firstoftwo}{%
 \@ifnum{\z@=\pdfoutput}{\@firstoftwo}{\@secondoftwo}%
}{%
 \providecommand\@@startlink[1]{\leavevmode}%
 \providecommand\@@endlink[0]{}%
}{%
 \providecommand\@@startlink[1]{%
  \leavevmode
  \pdfstartlink
   attr{/Border[0 0 1 ]/H/I/C[0 1 1]}%
   user{/Subtype/Link/A<</Type/Action/S/URI/URI(#1)>>}%
  \relax
 }%
 \providecommand\@@endlink[0]{\pdfendlink}%
}%
\providecommand \url  [0]{\begingroup\@sanitize \@url }%
\providecommand \@url [1]{\endgroup\@href {#1}{\urlprefix}}%
\providecommand \urlprefix [0]{URL }%
\providecommand \Eprint[0]{\href }%
\@ifxundefined \urlstyle {%
  \providecommand \doi [1]{doi:\discretionary{}{}{}#1}%
}{%
  \providecommand \doi [0]{doi:\discretionary{}{}{}\begingroup
  \urlstyle{rm}\Url }%
}%
\providecommand \doibase [0]{http://dx.doi.org/}%
\providecommand \Doi[1]{\href{\doibase#1}}%
\providecommand \bibAnnote [3]{%
  \BibitemShut{#1}%
  \begin{quotation}\noindent
    \textsc{Key:}\ #2\\\textsc{Annotation:}\ #3%
  \end{quotation}%
}%
\providecommand \bibAnnoteFile [2]{%
  \IfFileExists{#2}{\bibAnnote {#1} {#2} {\input{#2}}}{}%
}%
\providecommand \typeout [0]{\immediate \write \m@ne }%
\providecommand \selectlanguage [0]{\@gobble}%
\providecommand \bibinfo [0]{\@secondoftwo}%
\providecommand \bibfield [0]{\@secondoftwo}%
\providecommand \translation [1]{[#1]}%
\providecommand \BibitemOpen[0]{}%
\providecommand \bibitemStop [0]{}%
\providecommand \bibitemNoStop [0]{.\EOS\space}%
\providecommand \EOS [0]{\spacefactor3000\relax}%
\providecommand \BibitemShut [1]{\csname bibitem#1\endcsname}%
\bibitem{Veprek:Hard:99}%
  \BibitemOpen
  \bibfield{author}{%
  \bibinfo {author} {\bibfnamefont{S.}~\bibnamefont{Vep\v{r}ek}},\ }%
  \bibfield{journal}{%
  \bibinfo {journal} {Journal of Vacuum Science and Technology A}\ }%
  \textbf{\bibinfo {volume} {17}} (\bibinfo {year} {1999})%
  \bibAnnoteFile{NoStop}{Veprek:Hard:99}%
\bibitem{PRL:Jhi:Vacancy:TMNs:01}%
  \BibitemOpen
  \bibfield{author}{%
  \bibinfo {author} {\bibfnamefont{S.-H.}\ \bibnamefont{Jhi}}, \bibinfo
  {author} {\bibfnamefont{S.~G.}\ \bibnamefont{Louie}}, \bibinfo {author}
  {\bibfnamefont{M.~L.}\ \bibnamefont{Cohen}},\ and\ \bibinfo {author}
  {\bibfnamefont{J.}~\bibnamefont{Ihm}},\ }%
  \bibfield{journal}{%
  \Doi{10.1103/PhysRevLett.86.3348}{\bibinfo {journal} {Phys. Rev. Lett.}}\ }%
  \textbf{\bibinfo {volume} {86}},\ \bibinfo {pages} {3348} (\bibinfo {month}
  {Apr}\ \bibinfo {year} {2001}),\
  \url{http://link.aps.org/doi/10.1103/PhysRevLett.86.3348}%
  \bibAnnoteFile{NoStop}{PRL:Jhi:Vacancy:TMNs:01}%
\bibitem{Nature:Jhi:TMNs}%
  \BibitemOpen
  \bibfield{author}{%
  \bibinfo {author} {\bibfnamefont{S.-H.}\ \bibnamefont{Jhi}}, \bibinfo
  {author} {\bibfnamefont{J.}~\bibnamefont{Ihm}}, \bibinfo {author}
  {\bibfnamefont{S.~G.}\ \bibnamefont{Louie}},\ and\ \bibinfo {author}
  {\bibfnamefont{M.~L.}\ \bibnamefont{Cohen}},\ }%
  \bibfield{journal}{%
  \bibinfo {journal} {Nature}\ }%
  \textbf{\bibinfo {volume} {399}},\ \bibinfo {pages} {132} (\bibinfo {year}
  {1999})%
  \bibAnnoteFile{NoStop}{Nature:Jhi:TMNs}%
\bibitem{Hao:PRL:superhard}%
  \BibitemOpen
  \bibfield{author}{%
  \bibinfo {author} {\bibfnamefont{S.}~\bibnamefont{Hao}}, \bibinfo {author}
  {\bibfnamefont{B.}~\bibnamefont{Delley}}, \bibinfo {author}
  {\bibfnamefont{S.}~\bibnamefont{Veprek}},\ and\ \bibinfo {author}
  {\bibfnamefont{C.}~\bibnamefont{Stampfl}},\ }%
  \bibfield{journal}{%
  \Doi{10.1103/PhysRevLett.97.086102}{\bibinfo {journal} {Phys. Rev. Lett.}}\
  }%
  \textbf{\bibinfo {volume} {97}},\ \bibinfo {pages} {086102} (\bibinfo {month}
  {Aug}\ \bibinfo {year} {2006}),\
  \url{http://link.aps.org/doi/10.1103/PhysRevLett.97.086102}%
  \bibAnnoteFile{NoStop}{Hao:PRL:superhard}%
\bibitem{TMNs:superconductivity}%
  \BibitemOpen
  \bibfield{author}{%
  \bibinfo {author} {\bibfnamefont{S.}~\bibnamefont{Yamanaka}}, \bibinfo
  {author} {\bibfnamefont{K.-i.}\ \bibnamefont{Hotehama}},\ and\ \bibinfo
  {author} {\bibfnamefont{H.}~\bibnamefont{Kawaji}},\ }%
  \bibfield{journal}{%
  \bibinfo {journal} {Nature}\ }%
  \textbf{\bibinfo {volume} {392}},\ \bibinfo {pages} {580} (\bibinfo {year}
  {1998})%
  \bibAnnoteFile{NoStop}{TMNs:superconductivity}%
\bibitem{PRB:Steneteg:13:TMNs}%
  \BibitemOpen
  \bibfield{author}{%
  \bibinfo {author} {\bibfnamefont{P.}~\bibnamefont{Steneteg}}, \bibinfo
  {author} {\bibfnamefont{O.}~\bibnamefont{Hellman}}, \bibinfo {author}
  {\bibfnamefont{O.~Y.}\ \bibnamefont{Vekilova}}, \bibinfo {author}
  {\bibfnamefont{N.}~\bibnamefont{Shulumba}}, \bibinfo {author}
  {\bibfnamefont{F.}~\bibnamefont{Tasn\'adi}},\ and\ \bibinfo {author}
  {\bibfnamefont{I.~A.}\ \bibnamefont{Abrikosov}},\ }%
  \bibfield{journal}{%
  \Doi{10.1103/PhysRevB.87.094114}{\bibinfo {journal} {Phys. Rev. B}}\ }%
  \textbf{\bibinfo {volume} {87}},\ \bibinfo {pages} {094114} (\bibinfo {month}
  {Mar}\ \bibinfo {year} {2013}),\
  \url{http://link.aps.org/doi/10.1103/PhysRevB.87.094114}%
  \bibAnnoteFile{NoStop}{PRB:Steneteg:13:TMNs}%
\bibitem{Science:Sproul:hard}%
  \BibitemOpen
  \bibfield{author}{%
  \bibinfo {author} {\bibfnamefont{W.~D.}\ \bibnamefont{Sproul}},\ }%
  \bibfield{journal}{%
  \Doi{10.1126/science.273.5277.889}{\bibinfo {journal} {Science}}\ }%
  \textbf{\bibinfo {volume} {273}},\ \bibinfo {pages} {889} (\bibinfo {year}
  {1996}),\
  \Eprint{http://arxiv.org/abs/http://www.sciencemag.org/content/273/5277/889.full.pdf}{http://www.sciencemag.org/content/273/5277/889.full.pdf},\
  \url{http://www.sciencemag.org/content/273/5277/889.abstract}%
  \bibAnnoteFile{NoStop}{Science:Sproul:hard}%
\bibitem{PRB:1993:Haguland}%
  \BibitemOpen
  \bibfield{author}{%
  \bibinfo {author} {\bibfnamefont{J.}~\bibnamefont{H\"aglund}}, \bibinfo
  {author} {\bibfnamefont{A.}~\bibnamefont{Fern\'andez~Guillermet}}, \bibinfo
  {author} {\bibfnamefont{G.}~\bibnamefont{Grimvall}},\ and\ \bibinfo {author}
  {\bibfnamefont{M.}~\bibnamefont{K\"orling}},\ }%
  \bibfield{journal}{%
  \Doi{10.1103/PhysRevB.48.11685}{\bibinfo {journal} {Phys. Rev. B}}\ }%
  \textbf{\bibinfo {volume} {48}},\ \bibinfo {pages} {11685} (\bibinfo {month}
  {Oct}\ \bibinfo {year} {1993}),\
  \url{http://link.aps.org/doi/10.1103/PhysRevB.48.11685}%
  \bibAnnoteFile{NoStop}{PRB:1993:Haguland}%
\bibitem{Hultman:2000}%
  \BibitemOpen
  \bibfield{author}{%
  \bibinfo {author} {\bibfnamefont{L.}~\bibnamefont{Hultman}},\ }%
  \bibfield{journal}{%
  \Doi{http://dx.doi.org/10.1016/S0042-207X(00)00143-3}{\bibinfo {journal}
  {Vacuum}}\ }%
  \textbf{\bibinfo {volume} {57}},\ \bibinfo {pages} {1 } (\bibinfo {year}
  {2000}),\ ISSN \bibinfo {issn} {0042-207X},\
  \url{http://www.sciencedirect.com/science/article/pii/S0042207X00001433}%
  \bibAnnoteFile{NoStop}{Hultman:2000}%
\bibitem{Zhang:TMNs:2003:SCT}%
  \BibitemOpen
  \bibfield{author}{%
  \bibinfo {author} {\bibfnamefont{S.}~\bibnamefont{Zhang}}, \bibinfo {author}
  {\bibfnamefont{D.}~\bibnamefont{Sun}}, \bibinfo {author}
  {\bibfnamefont{Y.}~\bibnamefont{Fu}},\ and\ \bibinfo {author}
  {\bibfnamefont{H.}~\bibnamefont{Du}},\ }%
  \bibfield{journal}{%
  \Doi{http://dx.doi.org/10.1016/S0257-8972(02)00903-9}{\bibinfo {journal}
  {Surface and Coatings Technology}}\ }%
  \textbf{\bibinfo {volume} {167}},\ \bibinfo {pages} {113 } (\bibinfo {year}
  {2003}),\ ISSN \bibinfo {issn} {0257-8972},\ \bibinfo {note} {proceedings of
  the Symposium on Technological Advances and Performance of Engineering Thin
  Films and Surface Coatings at the 1st International Conference on Materials
  Processing for Properties and Performance (MP3)},\
  \url{http://www.sciencedirect.com/science/article/pii/S0257897202009039}%
  \bibAnnoteFile{NoStop}{Zhang:TMNs:2003:SCT}%
\bibitem{TiNZrNCrN:SCT:1998}%
  \BibitemOpen
  \bibfield{author}{%
  \bibinfo {author} {\bibfnamefont{I.}~\bibnamefont{Milo\v{s}ev}}, \bibinfo
  {author} {\bibfnamefont{H.-H.}\ \bibnamefont{Strehblow}},\ and\ \bibinfo
  {author} {\bibfnamefont{B.}~\bibnamefont{Navin\v{s}ek}},\ }%
  \bibfield{journal}{%
  \Doi{http://dx.doi.org/10.1016/S0040-6090(97)00069-2}{\bibinfo {journal}
  {Thin Solid Films}}\ }%
  \textbf{\bibinfo {volume} {303}},\ \bibinfo {pages} {246 } (\bibinfo {year}
  {1997}),\ ISSN \bibinfo {issn} {0040-6090},\
  \url{http://www.sciencedirect.com/science/article/pii/S0040609097000692}%
  \bibAnnoteFile{NoStop}{TiNZrNCrN:SCT:1998}%
\bibitem{PRB:Tsetseris:Ndefects:07}%
  \BibitemOpen
  \bibfield{author}{%
  \bibinfo {author} {\bibfnamefont{L.}~\bibnamefont{Tsetseris}}, \bibinfo
  {author} {\bibfnamefont{N.}~\bibnamefont{Kalfagiannis}}, \bibinfo {author}
  {\bibfnamefont{S.}~\bibnamefont{Logothetidis}},\ and\ \bibinfo {author}
  {\bibfnamefont{S.~T.}\ \bibnamefont{Pantelides}},\ }%
  \bibfield{journal}{%
  \Doi{10.1103/PhysRevB.76.224107}{\bibinfo {journal} {Phys. Rev. B}}\ }%
  \textbf{\bibinfo {volume} {76}},\ \bibinfo {pages} {224107} (\bibinfo {month}
  {Dec}\ \bibinfo {year} {2007}),\
  \url{http://link.aps.org/doi/10.1103/PhysRevB.76.224107}%
  \bibAnnoteFile{NoStop}{PRB:Tsetseris:Ndefects:07}%
\bibitem{PRL:TMNs:07:Ndefects}%
  \BibitemOpen
  \bibfield{author}{%
  \bibinfo {author} {\bibfnamefont{L.}~\bibnamefont{Tsetseris}}, \bibinfo
  {author} {\bibfnamefont{N.}~\bibnamefont{Kalfagiannis}}, \bibinfo {author}
  {\bibfnamefont{S.}~\bibnamefont{Logothetidis}},\ and\ \bibinfo {author}
  {\bibfnamefont{S.~T.}\ \bibnamefont{Pantelides}},\ }%
  \bibfield{journal}{%
  \Doi{10.1103/PhysRevLett.99.125503}{\bibinfo {journal} {Phys. Rev. Lett.}}\
  }%
  \textbf{\bibinfo {volume} {99}},\ \bibinfo {pages} {125503} (\bibinfo {month}
  {Sep}\ \bibinfo {year} {2007}),\
  \url{http://link.aps.org/doi/10.1103/PhysRevLett.99.125503}%
  \bibAnnoteFile{NoStop}{PRL:TMNs:07:Ndefects}%
\bibitem{MG:JAC:2011}%
  \BibitemOpen
  \bibfield{author}{%
  \bibinfo {author} {\bibfnamefont{M.}~\bibnamefont{Gupta}}, \bibinfo {author}
  {\bibfnamefont{A.}~\bibnamefont{Tayal}}, \bibinfo {author}
  {\bibfnamefont{A.}~\bibnamefont{Gupta}}, \bibinfo {author}
  {\bibfnamefont{V.}~\bibnamefont{Reddy}}, \bibinfo {author}
  {\bibfnamefont{M.}~\bibnamefont{Horisberger}},\ and\ \bibinfo {author}
  {\bibfnamefont{J.}~\bibnamefont{Stahn}},\ }%
  \bibfield{journal}{%
  \Doi{10.1016/j.jallcom.2011.04.139}{\bibinfo {journal} {J. Alloys and
  Compounds}}\ }%
  \textbf{\bibinfo {volume} {509}},\ \bibinfo {pages} {8283 } (\bibinfo {year}
  {2011}),\ ISSN \bibinfo {issn} {0925-8388},\
  \url{http://www.sciencedirect.com/science/article/pii/S0925838811010206}%
  \bibAnnoteFile{NoStop}{MG:JAC:2011}%
\bibitem{gupta:JAP2011}%
  \BibitemOpen
  \bibfield{author}{%
  \bibinfo {author} {\bibfnamefont{M.}~\bibnamefont{Gupta}}, \bibinfo {author}
  {\bibfnamefont{A.}~\bibnamefont{Tayal}}, \bibinfo {author}
  {\bibfnamefont{A.}~\bibnamefont{Gupta}}, \bibinfo {author}
  {\bibfnamefont{R.}~\bibnamefont{Gupta}}, \bibinfo {author}
  {\bibfnamefont{J.}~\bibnamefont{Stahn}}, \bibinfo {author}
  {\bibfnamefont{M.}~\bibnamefont{Horisberger}},\ and\ \bibinfo {author}
  {\bibfnamefont{A.}~\bibnamefont{Wildes}},\ }%
  \bibfield{journal}{%
  \Doi{10.1063/1.3671532}{\bibinfo {journal} {J. Appl. Phys.}}\ }%
  \textbf{\bibinfo {volume} {110}},\ \bibinfo {eid} {123518} (\bibinfo {year}
  {2011})%
  \bibAnnoteFile{NoStop}{gupta:JAP2011}%
\bibitem{FeN:PRB:Houari}%
  \BibitemOpen
  \bibfield{author}{%
  \bibinfo {author} {\bibfnamefont{A.}~\bibnamefont{Houari}}, \bibinfo {author}
  {\bibfnamefont{S.~F.}\ \bibnamefont{Matar}}, \bibinfo {author}
  {\bibfnamefont{M.~A.}\ \bibnamefont{Belkhir}},\ and\ \bibinfo {author}
  {\bibfnamefont{M.}~\bibnamefont{Nakhl}},\ }%
  \bibfield{journal}{%
  \Doi{10.1103/PhysRevB.75.064420}{\bibinfo {journal} {Phys. Rev. B}}\ }%
  \textbf{\bibinfo {volume} {75}},\ \bibinfo {pages} {064420} (\bibinfo {month}
  {Feb}\ \bibinfo {year} {2007}),\
  \url{http://link.aps.org/doi/10.1103/PhysRevB.75.064420}%
  \bibAnnoteFile{NoStop}{FeN:PRB:Houari}%
\bibitem{Jouanny2010TSF}%
  \BibitemOpen
  \bibfield{author}{%
  \bibinfo {author} {\bibfnamefont{I.}~\bibnamefont{Jouanny}}, \bibinfo
  {author} {\bibfnamefont{P.}~\bibnamefont{Weisbecker}}, \bibinfo {author}
  {\bibfnamefont{V.}~\bibnamefont{Demange}}, \bibinfo {author}
  {\bibfnamefont{M.}~\bibnamefont{Grafout\'{e}}}, \bibinfo {author}
  {\bibfnamefont{O.}~\bibnamefont{Pe{\~{n}}a}},\ and\ \bibinfo {author}
  {\bibfnamefont{E.}~\bibnamefont{Bauer-Grosse}},\ }%
  \bibfield{journal}{%
  \Doi{10.1016/j.tsf.2009.07.039}{\bibinfo {journal} {Thin Solid Films}}\ }%
  \textbf{\bibinfo {volume} {518}},\ \bibinfo {pages} {1883 } (\bibinfo {year}
  {2010})%
  \bibAnnoteFile{NoStop}{Jouanny2010TSF}%
\bibitem{Liu:CoN:14:JAC}%
  \BibitemOpen
  \bibfield{author}{%
  \bibinfo {author} {\bibfnamefont{X.}~\bibnamefont{Liu}}, \bibinfo {author}
  {\bibfnamefont{H.}~\bibnamefont{Lu}}, \bibinfo {author}
  {\bibfnamefont{M.}~\bibnamefont{He}}, \bibinfo {author}
  {\bibfnamefont{K.}~\bibnamefont{Jin}}, \bibinfo {author}
  {\bibfnamefont{G.}~\bibnamefont{Yang}}, \bibinfo {author}
  {\bibfnamefont{H.}~\bibnamefont{Ni}},\ and\ \bibinfo {author}
  {\bibfnamefont{K.}~\bibnamefont{Zhao}},\ }%
  \bibfield{journal}{%
  \Doi{http://dx.doi.org/10.1016/j.jallcom.2013.08.001}{\bibinfo {journal}
  {Journal of Alloys and Compounds}}\ }%
  \textbf{\bibinfo {volume} {582}},\ \bibinfo {pages} {75 } (\bibinfo {year}
  {2014}),\ ISSN \bibinfo {issn} {0925-8388},\
  \url{http://www.sciencedirect.com/science/article/pii/S092583881301846X}%
  \bibAnnoteFile{NoStop}{Liu:CoN:14:JAC}%
\bibitem{Navio.PRB08}%
  \BibitemOpen
  \bibfield{author}{%
  \bibinfo {author} {\bibfnamefont{C.}~\bibnamefont{Nav\'{\i}o}}, \bibinfo
  {author} {\bibfnamefont{J.}~\bibnamefont{Alvarez}}, \bibinfo {author}
  {\bibfnamefont{M.~J.}\ \bibnamefont{Capitan}}, \bibinfo {author}
  {\bibfnamefont{F.}~\bibnamefont{Yndurain}},\ and\ \bibinfo {author}
  {\bibfnamefont{R.}~\bibnamefont{Miranda}},\ }%
  \bibfield{journal}{%
  \Doi{doi:10.1103/PhysRevB.78.155417}{\bibinfo {journal} {Phys. Rev. B}}\ }%
  \textbf{\bibinfo {volume} {78}},\ \bibinfo {pages} {155417} (\bibinfo {year}
  {2008})%
  \bibAnnoteFile{NoStop}{Navio.PRB08}%
\bibitem{Wang:CoN:TSF:09}%
  \BibitemOpen
  \bibfield{author}{%
  \bibinfo {author} {\bibfnamefont{X.}~\bibnamefont{Wang}}, \bibinfo {author}
  {\bibfnamefont{H.}~\bibnamefont{Jia}}, \bibinfo {author}
  {\bibfnamefont{W.}~\bibnamefont{Zheng}}, \bibinfo {author}
  {\bibfnamefont{Y.}~\bibnamefont{Chen}},\ and\ \bibinfo {author}
  {\bibfnamefont{S.}~\bibnamefont{Feng}},\ }%
  \bibfield{journal}{%
  \Doi{http://dx.doi.org/10.1016/j.tsf.2009.03.171}{\bibinfo {journal} {Thin
  Solid Films}}\ }%
  \textbf{\bibinfo {volume} {517}},\ \bibinfo {pages} {4419 } (\bibinfo {year}
  {2009}),\ ISSN \bibinfo {issn} {0040-6090},\
  \url{http://www.sciencedirect.com/science/article/pii/S0040609009006695}%
  \bibAnnoteFile{NoStop}{Wang:CoN:TSF:09}%
\bibitem{Bhattacharyya:FeN:Review}%
  \BibitemOpen
  \bibfield{author}{%
  \bibinfo {author} {\bibfnamefont{S.}~\bibnamefont{Bhattacharyya}},\ }%
  \bibfield{journal}{%
  \Doi{10.1021/jp510606z}{\bibinfo {journal} {The Journal of Physical Chemistry
  C}}\ }%
  \textbf{\bibinfo {volume} {119}},\ \bibinfo {pages} {1601} (\bibinfo {year}
  {2015}),\
  \Eprint{http://arxiv.org/abs/http://dx.doi.org/10.1021/jp510606z}{http://dx.doi.org/10.1021/jp510606z},\
  \url{http://dx.doi.org/10.1021/jp510606z}%
  \bibAnnoteFile{NoStop}{Bhattacharyya:FeN:Review}%
\bibitem{Schaaf.PMS.2002}%
  \BibitemOpen
  \bibfield{author}{%
  \bibinfo {author} {\bibfnamefont{P.}~\bibnamefont{Schaaf}},\ }%
  \bibfield{journal}{%
  \Doi{DOI: 10.1016/S0079-6425(00)00003-7}{\bibinfo {journal} {Prog. Mater.
  Sci.}}\ }%
  \textbf{\bibinfo {volume} {47}},\ \bibinfo {pages} {1 } (\bibinfo {year}
  {2002})%
  \bibAnnoteFile{NoStop}{Schaaf.PMS.2002}%
\bibitem{JVSTA:Fang:CoN}%
  \BibitemOpen
  \bibfield{author}{%
  \bibinfo {author} {\bibfnamefont{J.-S.}\ \bibnamefont{Fang}}, \bibinfo
  {author} {\bibfnamefont{L.-C.}\ \bibnamefont{Yang}}, \bibinfo {author}
  {\bibfnamefont{C.-S.}\ \bibnamefont{Hsu}}, \bibinfo {author}
  {\bibfnamefont{G.-S.}\ \bibnamefont{Chen}}, \bibinfo {author}
  {\bibfnamefont{Y.-W.}\ \bibnamefont{Lin}},\ and\ \bibinfo {author}
  {\bibfnamefont{G.-S.}\ \bibnamefont{Chen}},\ }%
  \bibfield{journal}{%
  \bibinfo {journal} {Journal of Vacuum Science \& Technology A}\ }%
  \textbf{\bibinfo {volume} {22}} (\bibinfo {year} {2004})%
  \bibAnnoteFile{NoStop}{JVSTA:Fang:CoN}%
\bibitem{Gupta_JAC01}%
  \BibitemOpen
  \bibfield{author}{%
  \bibinfo {author} {\bibfnamefont{M.}~\bibnamefont{Gupta}}, \bibinfo {author}
  {\bibfnamefont{A.}~\bibnamefont{Gupta}}, \bibinfo {author}
  {\bibfnamefont{P.}~\bibnamefont{Bhattacharya}}, \bibinfo {author}
  {\bibfnamefont{P.}~\bibnamefont{Misra}},\ and\ \bibinfo {author}
  {\bibfnamefont{L.}~\bibnamefont{Kukreja}},\ }%
  \bibfield{journal}{%
  \Doi{10.1016/S0925-8388(01)01316-0}{\bibinfo {journal} {J. Alloys and
  Compounds}}\ }%
  \textbf{\bibinfo {volume} {326}},\ \bibinfo {pages} {265 } (\bibinfo {year}
  {2001}),\
  \url{http://www.sciencedirect.com/science/article/pii/S0925838801013160}%
  \bibAnnoteFile{NoStop}{Gupta_JAC01}%
\bibitem{Naito:FeN:14}%
  \BibitemOpen
  \bibfield{author}{%
  \bibinfo {author} {\bibfnamefont{M.}~\bibnamefont{Naito}}, \bibinfo {author}
  {\bibfnamefont{K.}~\bibnamefont{Uehara}}, \bibinfo {author}
  {\bibfnamefont{R.}~\bibnamefont{Takeda}}, \bibinfo {author}
  {\bibfnamefont{Y.}~\bibnamefont{Taniyasu}},\ and\ \bibinfo {author}
  {\bibfnamefont{H.}~\bibnamefont{Yamamoto}},\ }%
  \bibfield{journal}{%
  \Doi{http://dx.doi.org/10.1016/j.jcrysgro.2014.12.022}{\bibinfo {journal}
  {Journal of Crystal Growth}}\ }%
  \textbf{\bibinfo {volume} {415}},\ \bibinfo {pages} {36 } (\bibinfo {year}
  {2015}),\ ISSN \bibinfo {issn} {0022-0248},\
  \url{http://www.sciencedirect.com/science/article/pii/S0022024814008379}%
  \bibAnnoteFile{NoStop}{Naito:FeN:14}%
\bibitem{Vempaire:Ni3N:JAP:09}%
  \BibitemOpen
  \bibfield{author}{%
  \bibinfo {author} {\bibfnamefont{D.}~\bibnamefont{Vempaire}}, \bibinfo
  {author} {\bibfnamefont{F.}~\bibnamefont{Fettar}}, \bibinfo {author}
  {\bibfnamefont{L.}~\bibnamefont{Ortega}}, \bibinfo {author}
  {\bibfnamefont{F.}~\bibnamefont{Pierre}}, \bibinfo {author}
  {\bibfnamefont{S.}~\bibnamefont{Miraglia}}, \bibinfo {author}
  {\bibfnamefont{A.}~\bibnamefont{Sulpice}}, \bibinfo {author}
  {\bibfnamefont{J.}~\bibnamefont{Pelletier}}, \bibinfo {author}
  {\bibfnamefont{E.~K.}\ \bibnamefont{Hlil}},\ and\ \bibinfo {author}
  {\bibfnamefont{D.}~\bibnamefont{Fruchart}},\ }%
  \bibfield{journal}{%
  \Doi{http://dx.doi.org/10.1063/1.3238290}{\bibinfo {journal} {Journal of
  Applied Physics}}\ }%
  \textbf{\bibinfo {volume} {106}},\ \bibinfo {eid} {073911} (\bibinfo {year}
  {2009}),\
  \url{http://scitation.aip.org/content/aip/journal/jap/106/7/10.1063/1.3238290}%
  \bibAnnoteFile{NoStop}{Vempaire:Ni3N:JAP:09}%
\bibitem{Nishihara:Ni2N:2014}%
  \BibitemOpen
  \bibfield{author}{%
  \bibinfo {author} {\bibfnamefont{H.}~\bibnamefont{Nishihara}}, \bibinfo
  {author} {\bibfnamefont{K.}~\bibnamefont{Suzuki}}, \bibinfo {author}
  {\bibfnamefont{R.}~\bibnamefont{Umetsu}}, \bibinfo {author}
  {\bibfnamefont{T.}~\bibnamefont{Kanomata}}, \bibinfo {author}
  {\bibfnamefont{T.}~\bibnamefont{Kaneko}}, \bibinfo {author}
  {\bibfnamefont{M.}~\bibnamefont{Zhou}}, \bibinfo {author}
  {\bibfnamefont{M.}~\bibnamefont{Tsujikawa}}, \bibinfo {author}
  {\bibfnamefont{M.}~\bibnamefont{Shirai}}, \bibinfo {author}
  {\bibfnamefont{T.}~\bibnamefont{Sakon}}, \bibinfo {author}
  {\bibfnamefont{T.}~\bibnamefont{Wada}}, \bibinfo {author}
  {\bibfnamefont{K.}~\bibnamefont{Terashima}},\ and\ \bibinfo {author}
  {\bibfnamefont{S.}~\bibnamefont{Imada}},\ }%
  \bibfield{journal}{%
  \Doi{http://dx.doi.org/10.1016/j.physb.2014.05.016}{\bibinfo {journal}
  {Physica B: Condensed Matter}}\ }%
  \textbf{\bibinfo {volume} {449}},\ \bibinfo {pages} {85 } (\bibinfo {year}
  {2014}),\ ISSN \bibinfo {issn} {0921-4526},\
  \url{http://www.sciencedirect.com/science/article/pii/S0921452614003998}%
  \bibAnnoteFile{NoStop}{Nishihara:Ni2N:2014}%
\bibitem{Navio:APL:2009}%
  \BibitemOpen
  \bibfield{author}{%
  \bibinfo {author} {\bibfnamefont{C.}~\bibnamefont{Nav\'{\i}o}}, \bibinfo
  {author} {\bibfnamefont{J.}~\bibnamefont{Alvarez}}, \bibinfo {author}
  {\bibfnamefont{M.~J.}\ \bibnamefont{Capitan}}, \bibinfo {author}
  {\bibfnamefont{J.}~\bibnamefont{Camarero}},\ and\ \bibinfo {author}
  {\bibfnamefont{R.}~\bibnamefont{Miranda}},\ }%
  \bibfield{journal}{%
  \Doi{doi:10.1063/1.3159630}{\bibinfo {journal} {Appl. Phys. Lett.}}\ }%
  \textbf{\bibinfo {volume} {94}},\ \bibinfo {pages} {263112} (\bibinfo {year}
  {2009})%
  \bibAnnoteFile{NoStop}{Navio:APL:2009}%
\bibitem{Navio.NJP2010}%
  \BibitemOpen
  \bibfield{author}{%
  \bibinfo {author} {\bibfnamefont{C.}~\bibnamefont{Nav\'{\i}o}}, \bibinfo
  {author} {\bibfnamefont{M.~J.}\ \bibnamefont{Capit\'{a}n}}, \bibinfo {author}
  {\bibfnamefont{J.}~\bibnamefont{\'{A}lvarez}}, \bibinfo {author}
  {\bibfnamefont{R.}~\bibnamefont{Miranda}},\ and\ \bibinfo {author}
  {\bibfnamefont{F.}~\bibnamefont{Yndurain}},\ }%
  \bibfield{journal}{%
  \bibinfo {journal} {New J. Phys.}\ }%
  \textbf{\bibinfo {volume} {12}},\ \bibinfo {pages} {073004} (\bibinfo {year}
  {2010})%
  \bibAnnoteFile{NoStop}{Navio.NJP2010}%
\bibitem{Faupel_RMP03}%
  \BibitemOpen
  \bibfield{author}{%
  \bibinfo {author} {\bibfnamefont{F.}~\bibnamefont{Faupel}}, \bibinfo {author}
  {\bibfnamefont{W.}~\bibnamefont{Frank}}, \bibinfo {author}
  {\bibfnamefont{M.~P.}\ \bibnamefont{Macht}}, \bibinfo {author}
  {\bibfnamefont{H.}~\bibnamefont{Mehrer}}, \bibinfo {author}
  {\bibfnamefont{K.}~\bibnamefont{R{\"{a}}tzke}}, \bibinfo {author}
  {\bibfnamefont{H.}~\bibnamefont{Schober}}, \bibinfo {author}
  {\bibfnamefont{S.~K.}\ \bibnamefont{Sharma}},\ and\ \bibinfo {author}
  {\bibfnamefont{H.}~\bibnamefont{Teichler}},\ }%
  \bibfield{journal}{%
  \bibinfo {journal} {Rev. Mod. Phys.}\ }%
  \textbf{\bibinfo {volume} {75}},\ \bibinfo {pages} {237} (\bibinfo {year}
  {2003})%
  \bibAnnoteFile{NoStop}{Faupel_RMP03}%
\bibitem{matzke1992diffusion}%
  \BibitemOpen
  \bibfield{author}{%
  \bibinfo {author} {\bibfnamefont{H.}~\bibnamefont{Matzke}},\ }%
  in\ \emph{\bibinfo {booktitle} {Defect and Diffusion Forum}},\ Vol.~\bibinfo
  {volume} {83}\ (\bibinfo {organization} {Trans Tech Publ},\ \bibinfo {year}
  {1992})\ pp.\ \bibinfo {pages} {111--130}%
  \bibAnnoteFile{NoStop}{matzke1992diffusion}%
\bibitem{PRB:AT:2014}%
  \BibitemOpen
  \bibfield{author}{%
  \bibinfo {author} {\bibfnamefont{A.}~\bibnamefont{Tayal}}, \bibinfo {author}
  {\bibfnamefont{M.}~\bibnamefont{Gupta}}, \bibinfo {author}
  {\bibfnamefont{N.~P.}\ \bibnamefont{Lalla}}, \bibinfo {author}
  {\bibfnamefont{A.}~\bibnamefont{Gupta}}, \bibinfo {author}
  {\bibfnamefont{M.}~\bibnamefont{Horisberger}}, \bibinfo {author}
  {\bibfnamefont{J.}~\bibnamefont{Stahn}}, \bibinfo {author}
  {\bibfnamefont{K.}~\bibnamefont{Schlage}},\ and\ \bibinfo {author}
  {\bibfnamefont{H.-C.}\ \bibnamefont{Wille}},\ }%
  \bibfield{journal}{%
  \Doi{10.1103/PhysRevB.90.144412}{\bibinfo {journal} {Phys. Rev. B}}\ }%
  \textbf{\bibinfo {volume} {90}},\ \bibinfo {pages} {144412} (\bibinfo {month}
  {Oct}\ \bibinfo {year} {2014}),\
  \url{http://link.aps.org/doi/10.1103/PhysRevB.90.144412}%
  \bibAnnoteFile{NoStop}{PRB:AT:2014}%
\bibitem{Harald.APL04}%
  \BibitemOpen
  \bibfield{author}{%
  \bibinfo {author} {\bibfnamefont{H.}~\bibnamefont{Schmidt}}, \bibinfo
  {author} {\bibfnamefont{G.}~\bibnamefont{Borchardt}}, \bibinfo {author}
  {\bibfnamefont{M.}~\bibnamefont{Rudolphi}}, \bibinfo {author}
  {\bibfnamefont{H.}~\bibnamefont{Baumann}},\ and\ \bibinfo {author}
  {\bibfnamefont{M.}~\bibnamefont{Bruns}},\ }%
  \bibfield{journal}{%
  \Doi{10.1063/1.1769594}{\bibinfo {journal} {Appl. Phys. Letters}}\ }%
  \textbf{\bibinfo {volume} {85}},\ \bibinfo {pages} {582} (\bibinfo {year}
  {2004})%
  \bibAnnoteFile{NoStop}{Harald.APL04}%
\bibitem{Harald.PRL06}%
  \BibitemOpen
  \bibfield{author}{%
  \bibinfo {author} {\bibfnamefont{H.}~\bibnamefont{Schmidt}}, \bibinfo
  {author} {\bibfnamefont{M.}~\bibnamefont{Gupta}},\ and\ \bibinfo {author}
  {\bibfnamefont{M.}~\bibnamefont{Bruns}},\ }%
  \bibfield{journal}{%
  \Doi{10.1103/PhysRevLett.96.055901}{\bibinfo {journal} {Phys. Rev. Lett.}}\
  }%
  \textbf{\bibinfo {volume} {96}},\ \bibinfo {pages} {055901} (\bibinfo {year}
  {2006})%
  \bibAnnoteFile{NoStop}{Harald.PRL06}%
\bibitem{LeClaire}%
  \BibitemOpen
  \bibfield{author}{%
  \bibinfo {author} {\bibfnamefont{A.~D.~L.}\ \bibnamefont{Claire}},\ }%
  \bibfield{journal}{%
  \bibinfo {journal} {British Journal of Applied Physics}\ }%
  \textbf{\bibinfo {volume} {14}},\ \bibinfo {pages} {351} (\bibinfo {year}
  {1963}),\ \url{http://stacks.iop.org/0508-3443/14/i=6/a=317}%
  \bibAnnoteFile{NoStop}{LeClaire}%
\bibitem{Book:Kaur1995}%
  \BibitemOpen
  \bibfield{author}{%
  \bibinfo {author} {\bibfnamefont{I.}~\bibnamefont{Kaur}}, \bibinfo {author}
  {\bibfnamefont{Y.}~\bibnamefont{Mishin}},\ and\ \bibinfo {author}
  {\bibfnamefont{W.}~\bibnamefont{Gust}},\ }%
  \emph{\bibinfo {title} {Fundamentals of Grain and Interphase Boundary
  Diffusion}}\ (\bibinfo {publisher} {Wiley},\ \bibinfo {year} {1995})\ ISBN
  \bibinfo {isbn} {9780471938194},\
  \url{http://books.google.co.in/books?id=RnmAQgAACAAJ}%
  \bibAnnoteFile{NoStop}{Book:Kaur1995}%
\bibitem{BL09:Indus2}%
  \BibitemOpen
  \bibfield{author}{%
  \bibinfo {author} {\bibfnamefont{S.}~\bibnamefont{Basu}}, \bibinfo {author}
  {\bibfnamefont{C.}~\bibnamefont{Nayak}}, \bibinfo {author}
  {\bibfnamefont{A.~K.}\ \bibnamefont{Yadav}}, \bibinfo {author}
  {\bibfnamefont{A.}~\bibnamefont{Agrawal}}, \bibinfo {author}
  {\bibfnamefont{A.~K.}\ \bibnamefont{Poswal}}, \bibinfo {author}
  {\bibfnamefont{D.}~\bibnamefont{Bhattacharyya}}, \bibinfo {author}
  {\bibfnamefont{S.~N.}\ \bibnamefont{Jha}},\ and\ \bibinfo {author}
  {\bibfnamefont{N.~K.}\ \bibnamefont{Sahoo}},\ }%
  \bibfield{journal}{%
  \bibinfo {journal} {Journal of Physics: Conference Series}\ }%
  \textbf{\bibinfo {volume} {493}},\ \bibinfo {pages} {012032} (\bibinfo {year}
  {2014}),\ \url{http://stacks.iop.org/1742-6596/493/i=1/a=012032}%
  \bibAnnoteFile{NoStop}{BL09:Indus2}%
\bibitem{Phase:SXAS:BL01}%
  \BibitemOpen
  \bibfield{author}{%
  \bibinfo {author} {\bibfnamefont{D.~M.}\ \bibnamefont{Phase}}, \bibinfo
  {author} {\bibfnamefont{M.}~\bibnamefont{Gupta}}, \bibinfo {author}
  {\bibfnamefont{S.}~\bibnamefont{Potdar}}, \bibinfo {author}
  {\bibfnamefont{L.}~\bibnamefont{Behera}}, \bibinfo {author}
  {\bibfnamefont{R.}~\bibnamefont{Sah}},\ and\ \bibinfo {author}
  {\bibfnamefont{A.}~\bibnamefont{Gupta}},\ }%
  \bibfield{journal}{%
  \Doi{http://dx.doi.org/10.1063/1.4872719}{\bibinfo {journal} {AIP Conference
  Proceedings}}\ }%
  \textbf{\bibinfo {volume} {1591}},\ \bibinfo {pages} {685} (\bibinfo {year}
  {2014}),\
  \url{http://scitation.aip.org/content/aip/proceeding/aipcp/10.1063/1.4872719}%
  \bibAnnoteFile{NoStop}{Phase:SXAS:BL01}%
\bibitem{Guinier_XRD}%
  \BibitemOpen
  \bibinfo {author} {\bibfnamefont{A.}~\bibnamefont{Guinier}}%
  \bibAnnoteFile{NoStop}{Guinier_XRD}%
\bibitem{warrenXRD}%
  \BibitemOpen
\bibfield{author}{%
    }%
  \bibfield{author}{%
  \bibinfo {author} {\bibfnamefont{B.~E.}\ \bibnamefont{Warren}},\ }%
  \emph{\bibinfo {title} {X-ray Diffraction}}\ (\bibinfo {publisher} {Courier
  Corporation},\ \bibinfo {year} {1969})%
  \bibAnnoteFile{NoStop}{warrenXRD}%
\bibitem{Cullity_XRD}%
  \BibitemOpen
  \bibfield{author}{%
  \bibinfo {author} {\bibfnamefont{B.~D.}\ \bibnamefont{Cullity}},\ }%
  \emph{\bibinfo {title} {Elements of X-ray Diffraction}}\ (\bibinfo
  {publisher} {Addison-Wesley,MA},\ \bibinfo {year} {1978})%
  \bibAnnoteFile{NoStop}{Cullity_XRD}%
\bibitem{Normos:brand:95}%
  \BibitemOpen
  \bibfield{author}{%
  \bibinfo {author} {\bibfnamefont{R.}~\bibnamefont{Brand}},\ }%
  \bibfield{journal}{%
  \bibinfo {journal} {Wissenschaftlich Elektronik GmbH, Starnberg}}%
   (\bibinfo {year} {1995})%
  \bibAnnoteFile{NoStop}{Normos:brand:95}%
\bibitem{Borsa.HI.2003}%
  \BibitemOpen
  \bibfield{author}{%
  \bibinfo {author} {\bibfnamefont{D.~M.}\ \bibnamefont{Borsa}}\ and\ \bibinfo
  {author} {\bibfnamefont{D.~O.}\ \bibnamefont{Boerma}},\ }%
  \bibfield{journal}{%
  \Doi{10.1023/B:HYPE.0000020403.64670.02}{\bibinfo {journal} {Hyp.\,Int.}}\ }%
  \textbf{\bibinfo {volume} {151-152}},\ \bibinfo {pages} {31} (\bibinfo {year}
  {2003})%
  \bibAnnoteFile{NoStop}{Borsa.HI.2003}%
\bibitem{JPC:chen:FexN:1983}%
  \BibitemOpen
  \bibfield{author}{%
  \bibinfo {author} {\bibfnamefont{G.}~\bibnamefont{Chen}}, \bibinfo {author}
  {\bibfnamefont{N.}~\bibnamefont{Jaggi}}, \bibinfo {author}
  {\bibfnamefont{J.}~\bibnamefont{Butt}}, \bibinfo {author}
  {\bibfnamefont{E.}~\bibnamefont{Yeh}},\ and\ \bibinfo {author}
  {\bibfnamefont{L.}~\bibnamefont{Schwartz}},\ }%
  \bibfield{journal}{%
  \bibinfo {journal} {Journal of physical chemistry}\ }%
  \textbf{\bibinfo {volume} {87}},\ \bibinfo {pages} {5326} (\bibinfo {year}
  {1983})%
  \bibAnnoteFile{NoStop}{JPC:chen:FexN:1983}%
\bibitem{Bunker:XAFS}%
  \BibitemOpen
  \bibfield{author}{%
  \bibinfo {author} {\bibfnamefont{G.}~\bibnamefont{Bunker}},\ }%
  \emph{\bibinfo {title} {Introduction to XAFS: a practical guide to X-ray
  absorption fine structure spectroscopy}}\ (\bibinfo {publisher} {Cambridge
  University Press},\ \bibinfo {year} {2010})%
  \bibAnnoteFile{NoStop}{Bunker:XAFS}%
\bibitem{Pre-edge:XAS:Yamamoto}%
  \BibitemOpen
  \bibfield{author}{%
  \bibinfo {author} {\bibfnamefont{T.}~\bibnamefont{Yamamoto}},\ }%
  \bibfield{journal}{%
  \Doi{10.1002/xrs.1103}{\bibinfo {journal} {X-Ray Spectrometry}}\ }%
  \textbf{\bibinfo {volume} {37}},\ \bibinfo {pages} {572} (\bibinfo {year}
  {2008}),\ ISSN \bibinfo {issn} {1097-4539},\
  \url{http://dx.doi.org/10.1002/xrs.1103}%
  \bibAnnoteFile{NoStop}{Pre-edge:XAS:Yamamoto}%
\bibitem{Pre-edge:XAS:Cr}%
  \BibitemOpen
  \bibfield{author}{%
  \bibinfo {author} {\bibfnamefont{A.}~\bibnamefont{Pantelouris}}, \bibinfo
  {author} {\bibfnamefont{H.}~\bibnamefont{Modrow}}, \bibinfo {author}
  {\bibfnamefont{M.}~\bibnamefont{Pantelouris}}, \bibinfo {author}
  {\bibfnamefont{J.}~\bibnamefont{Hormes}},\ and\ \bibinfo {author}
  {\bibfnamefont{D.}~\bibnamefont{Reinen}},\ }%
  \bibfield{journal}{%
  \Doi{http://dx.doi.org/10.1016/j.chemphys.2003.12.017}{\bibinfo {journal}
  {Chemical Physics}}\ }%
  \textbf{\bibinfo {volume} {300}},\ \bibinfo {pages} {13 } (\bibinfo {year}
  {2004}),\ ISSN \bibinfo {issn} {0301-0104},\
  \url{http://www.sciencedirect.com/science/article/pii/S0301010404000059}%
  \bibAnnoteFile{NoStop}{Pre-edge:XAS:Cr}%
\bibitem{SSC:Mitterbauer:EELS:04}%
  \BibitemOpen
  \bibfield{author}{%
  \bibinfo {author} {\bibfnamefont{C.}~\bibnamefont{Mitterbauer}}, \bibinfo
  {author} {\bibfnamefont{C.}~\bibnamefont{HAbert}}, \bibinfo {author}
  {\bibfnamefont{G.}~\bibnamefont{Kothleitner}}, \bibinfo {author}
  {\bibfnamefont{F.}~\bibnamefont{Hofer}}, \bibinfo {author}
  {\bibfnamefont{P.}~\bibnamefont{Schattschneider}},\ and\ \bibinfo {author}
  {\bibfnamefont{H.}~\bibnamefont{Zandbergen}},\ }%
  \bibfield{journal}{%
  \Doi{http://dx.doi.org/10.1016/j.ssc.2004.01.045}{\bibinfo {journal} {Solid
  State Communications}}\ }%
  \textbf{\bibinfo {volume} {130}},\ \bibinfo {pages} {209 } (\bibinfo {year}
  {2004}),\ ISSN \bibinfo {issn} {0038-1098},\
  \url{http://www.sciencedirect.com/science/article/pii/S0038109804000936}%
  \bibAnnoteFile{NoStop}{SSC:Mitterbauer:EELS:04}%
\bibitem{Pfluger:EELS:89}%
  \BibitemOpen
  \bibfield{author}{%
  \bibinfo {author} {\bibfnamefont{J.}~\bibnamefont{Pfluger}}, \bibinfo
  {author} {\bibfnamefont{J.}~\bibnamefont{Fink}}, \bibinfo {author}
  {\bibfnamefont{G.}~\bibnamefont{Crecelius}}, \bibinfo {author}
  {\bibfnamefont{K.}~\bibnamefont{Bohnen}},\ and\ \bibinfo {author}
  {\bibfnamefont{H.}~\bibnamefont{Winter}},\ }%
  \bibfield{journal}{%
  \Doi{http://dx.doi.org/10.1016/0038-1098(82)90130-2}{\bibinfo {journal}
  {Solid State Communications}}\ }%
  \textbf{\bibinfo {volume} {44}},\ \bibinfo {pages} {489 } (\bibinfo {year}
  {1982}),\ ISSN \bibinfo {issn} {0038-1098},\
  \url{http://www.sciencedirect.com/science/article/pii/0038109882901302}%
  \bibAnnoteFile{NoStop}{Pfluger:EELS:89}%
\bibitem{Spapen:APL80}%
  \BibitemOpen
  \bibfield{author}{%
  \bibinfo {author} {\bibfnamefont{M.~P.}\ \bibnamefont{Rosenblum}}, \bibinfo
  {author} {\bibfnamefont{F.}~\bibnamefont{Spaepen}},\ and\ \bibinfo {author}
  {\bibfnamefont{D.}~\bibnamefont{Turnbull}},\ }%
  \bibfield{journal}{%
  \Doi{doi:10.1063/1.91818}{\bibinfo {journal} {Appl. Phys. Lett.}}\ }%
  \textbf{\bibinfo {volume} {37}},\ \bibinfo {pages} {184} (\bibinfo {year}
  {1980})%
  \bibAnnoteFile{NoStop}{Spapen:APL80}%
\bibitem{Greer-JMMM96}%
  \BibitemOpen
  \bibfield{author}{%
  \bibinfo {author} {\bibfnamefont{J.}~\bibnamefont{Speakman}}, \bibinfo
  {author} {\bibfnamefont{P.}~\bibnamefont{Rose}}, \bibinfo {author}
  {\bibfnamefont{J.}~\bibnamefont{Hunt}}, \bibinfo {author}
  {\bibfnamefont{N.}~\bibnamefont{Cowlam}}, \bibinfo {author}
  {\bibfnamefont{R.~E.}\ \bibnamefont{Somekh}},\ and\ \bibinfo {author}
  {\bibfnamefont{A.}~\bibnamefont{Greer}},\ }%
  \bibfield{journal}{%
  \bibinfo {journal} {J. Magn. Magn. Mat.}\ }%
  \textbf{\bibinfo {volume} {156}},\ \bibinfo {pages} {411} (\bibinfo {year}
  {1996})%
  \bibAnnoteFile{NoStop}{Greer-JMMM96}%
\bibitem{gupta:PRB04}%
  \BibitemOpen
  \bibfield{author}{%
  \bibinfo {author} {\bibfnamefont{M.}~\bibnamefont{Gupta}}, \bibinfo {author}
  {\bibfnamefont{A.}~\bibnamefont{Gupta}}, \bibinfo {author}
  {\bibfnamefont{J.}~\bibnamefont{Stahn}}, \bibinfo {author}
  {\bibfnamefont{M.}~\bibnamefont{Horisberger}}, \bibinfo {author}
  {\bibfnamefont{T.}~\bibnamefont{Gutberlet}},\ and\ \bibinfo {author}
  {\bibfnamefont{P.}~\bibnamefont{Allenspach}},\ }%
  \bibfield{journal}{%
  \Doi{doi:10.1103/PhysRevB.70.184206}{\bibinfo {journal} {Phys. Rev. B}}\ }%
  \textbf{\bibinfo {volume} {70}},\ \bibinfo {eid} {184206} (\bibinfo {year}
  {2004})%
  \bibAnnoteFile{NoStop}{gupta:PRB04}%
\bibitem{Parratt.PR54}%
  \BibitemOpen
  \bibfield{author}{%
  \bibinfo {author} {\bibfnamefont{L.~G.}\ \bibnamefont{Parratt}},\ }%
  \bibfield{journal}{%
  \bibinfo {journal} {Phys. Rev.}\ }%
  \textbf{\bibinfo {volume} {95}},\ \bibinfo {pages} {359} (\bibinfo {year}
  {1954})%
  \bibAnnoteFile{NoStop}{Parratt.PR54}%
\bibitem{Parratt32}%
  \BibitemOpen
  \bibfield{author}{%
  \bibinfo {author} {\bibfnamefont{C.}~\bibnamefont{Braun}},\ }%
  \emph{\bibinfo {title} {Parratt32- The Reflectivity Tool}}\ (\bibinfo
  {publisher} {HMI Berlin},\ \bibinfo {year} {1997-99})%
  \bibAnnoteFile{NoStop}{Parratt32}%
\bibitem{gupta:PRB02}%
  \BibitemOpen
  \bibfield{author}{%
  \bibinfo {author} {\bibfnamefont{M.}~\bibnamefont{Gupta}}, \bibinfo {author}
  {\bibfnamefont{A.}~\bibnamefont{Gupta}}, \bibinfo {author}
  {\bibfnamefont{S.}~\bibnamefont{Rajagopalan}},\ and\ \bibinfo {author}
  {\bibfnamefont{A.~K.}\ \bibnamefont{Tyagi}},\ }%
  \bibfield{journal}{%
  \Doi{doi:10.1103/PhysRevB.65.214204}{\bibinfo {journal} {Phys. Rev. B}}\ }%
  \textbf{\bibinfo {volume} {65}},\ \bibinfo {pages} {214204} (\bibinfo {year}
  {2002})%
  \bibAnnoteFile{NoStop}{gupta:PRB02}%
\bibitem{GBD:Mishin:1999}%
  \BibitemOpen
  \bibfield{author}{%
  \bibinfo {author} {\bibfnamefont{Y.}~\bibnamefont{Mishin}}\ and\ \bibinfo
  {author} {\bibfnamefont{C.}~\bibnamefont{Herzig}},\ }%
  \bibfield{journal}{%
  \Doi{http://dx.doi.org/10.1016/S0921-5093(98)00978-2}{\bibinfo {journal}
  {Materials Science and Engineering: A}}\ }%
  \textbf{\bibinfo {volume} {260}},\ \bibinfo {pages} {55 } (\bibinfo {year}
  {1999}),\ ISSN \bibinfo {issn} {0921-5093},\
  \url{http://www.sciencedirect.com/science/article/pii/S0921509398009782}%
  \bibAnnoteFile{NoStop}{GBD:Mishin:1999}%
\bibitem{Suzuoka:GBD}%
  \BibitemOpen
  \bibfield{author}{%
  \bibinfo {author} {\bibfnamefont{T.}~\bibnamefont{Suzuoka}},\ }%
  \bibfield{journal}{%
  \Doi{10.1143/JPSJ.19.839}{\bibinfo {journal} {Journal of the Physical Society
  of Japan}}\ }%
  \textbf{\bibinfo {volume} {19}},\ \bibinfo {pages} {839} (\bibinfo {year}
  {1964}),\
  \Eprint{http://arxiv.org/abs/http://dx.doi.org/10.1143/JPSJ.19.839}{http://dx.doi.org/10.1143/JPSJ.19.839},\
  \url{http://dx.doi.org/10.1143/JPSJ.19.839}%
  \bibAnnoteFile{NoStop}{Suzuoka:GBD}%
\bibitem{GB:Diffusion:Fisher}%
  \BibitemOpen
  \bibfield{author}{%
  \bibinfo {author} {\bibfnamefont{J.~C.}\ \bibnamefont{Fisher}},\ }%
  \bibfield{journal}{%
  \bibinfo {journal} {Journal of Applied Physics}\ }%
  \textbf{\bibinfo {volume} {22}},\ \bibinfo {pages} {74} (\bibinfo {year}
  {1951})%
  \bibAnnoteFile{NoStop}{GB:Diffusion:Fisher}%
\bibitem{Harrison}%
  \BibitemOpen
  \bibfield{author}{%
  \bibinfo {author} {\bibfnamefont{L.~G.}\ \bibnamefont{Harrison}},\ }%
  \bibfield{journal}{%
  \Doi{10.1039/TF9615701191}{\bibinfo {journal} {Trans. Faraday Soc.}}\ }%
  \textbf{\bibinfo {volume} {57}},\ \bibinfo {pages} {1191} (\bibinfo {year}
  {1961}),\ \url{http://dx.doi.org/10.1039/TF9615701191}%
  \bibAnnoteFile{NoStop}{Harrison}%
\bibitem{Mishin:95:TypeABC}%
  \BibitemOpen
  \bibfield{author}{%
  \bibinfo {author} {\bibfnamefont{Y.}~\bibnamefont{Mishin}}\ and\ \bibinfo
  {author} {\bibfnamefont{C.}~\bibnamefont{Herzig}},\ }%
  \bibfield{journal}{%
  \Doi{http://dx.doi.org/10.1016/0965-9773(95)00195-6}{\bibinfo {journal}
  {Nanostructured Materials}}\ }%
  \textbf{\bibinfo {volume} {6}},\ \bibinfo {pages} {859 } (\bibinfo {year}
  {1995}),\ ISSN \bibinfo {issn} {0965-9773},\ \bibinfo {note} {proceedings of
  the Second International Conference on Nanostructured Materials},\
  \url{http://www.sciencedirect.com/science/article/pii/0965977395001956}%
  \bibAnnoteFile{NoStop}{Mishin:95:TypeABC}%
\bibitem{Paduani:CuNCoNNiN:2008}%
  \BibitemOpen
  \bibfield{author}{%
  \bibinfo {author} {\bibfnamefont{C.}~\bibnamefont{Paduani}},\ }%
  \bibfield{journal}{%
  \Doi{http://dx.doi.org/10.1016/j.ssc.2008.09.010}{\bibinfo {journal} {Solid
  State Communications}}\ }%
  \textbf{\bibinfo {volume} {148}},\ \bibinfo {pages} {297 } (\bibinfo {year}
  {2008}),\ ISSN \bibinfo {issn} {0038-1098},\
  \url{http://www.sciencedirect.com/science/article/pii/S0038109808005164}%
  \bibAnnoteFile{NoStop}{Paduani:CuNCoNNiN:2008}%
\bibitem{Eck:JMC:99:TMNs}%
  \BibitemOpen
  \bibfield{author}{%
  \bibinfo {author} {\bibfnamefont{B.}~\bibnamefont{Eck}}, \bibinfo {author}
  {\bibfnamefont{R.}~\bibnamefont{Dronskowski}}, \bibinfo {author}
  {\bibfnamefont{M.}~\bibnamefont{Takahashi}},\ and\ \bibinfo {author}
  {\bibfnamefont{S.}~\bibnamefont{Kikkawa}},\ }%
  \bibfield{journal}{%
  \Doi{10.1039/A809935I}{\bibinfo {journal} {J. Mater. Chem.}}\ }%
  \textbf{\bibinfo {volume} {9}},\ \bibinfo {pages} {1527} (\bibinfo {year}
  {1999}),\ \url{http://dx.doi.org/10.1039/A809935I}%
  \bibAnnoteFile{NoStop}{Eck:JMC:99:TMNs}%
\end{thebibliography}

%

\end{document}